# Scale-by-scale kinetic energy flux calculations in simulations of rotating convection

**Youri H. Lemm[1], Xander M. de Wit[1], Rudie P.J. Kunnen[1]**

[1]Fluids and Flows group and J.M. Burgers Centre for Fluid Mechanics, Department of Applied Physics and Science Education, Eindhoven University of Technology, 5600 MB Eindhoven, The Netherlands

**Corresponding author:** Youri H. Lemm, y.h.lemm@tue.nl

Turbulence is an out-of-equilibrium flow state that is characterised by nonzero net fluxes of kinetic energy between different scales of the flow. These fluxes play a crucial role in the formation of characteristic flow structures in many turbulent flows encountered in nature. However, measuring these energy fluxes in practical settings can present a challenge in systems other than the case of unrestricted turbulence in an idealised periodic box. Here, we focus on rotating Rayleigh-Bénard convection, being the canonical model system to study geophysical and astrophysical flows. Owing to the effect of rotation, this flow can yield a split cascade, where part of the energy is transported to smaller scales (direct cascade), while another fraction is transported to larger scales (inverse cascade). We compare two different techniques for measuring these energy fluxes throughout the domain: one based on a spatial filtering approach and an adapted Fourier-based method. We show how one can use these methods to measure the energy flux adequately in the anisotropic, aperiodic domains encountered in rotating convection, even in domains with spatial confinement. Our measurements reveal that in the studied regime, the bulk flow is dominated by the direct cascade, while significant inverse cascading action is observed most strongly near the top and bottom plates, due to the vortex merging of Ekman plumes into larger flow structures.

## 1. Introduction

The behaviour of kinetic energy transport in turbulent flows is known to be dramatically different depending on the dimensionality of the system (Frisch 1995; Davidson 2015). In three-dimensional (3D) turbulence, energy injected at the largest scales is cascaded towards smaller and subsequently smaller scales. This cascading process ends with energy dissipating into heat on the smallest scales. This is referred to as the *direct* or *forward* cascade of energy (Pope 2000). When constrained to only two spatial dimensions, this cascading process is inverted, and the kinetic energy is transported mainly towards the large scales. This inversion of the energy cascade can result in an accumulation of kinetic energy at the largest scales. This phenomenon is known as the so-called *inverse* cascade of energy, which originates from the fact that this process is opposite to what is expected from unconstrained 3D turbulence. This inversion of the energy cascade is caused by the absence of vortex stretching in these two-dimensional (2D) turbulent flows (Fox & Davidson 2010; Boffetta & Ecke 2011).

This idealised system of two-dimensional turbulence is only rarely realized exactly, but many systems in our physical world do possess some degree of two-dimensionalisation, and hence we refer to these as cases of *quasi*-two-dimensional turbulence (Boffetta & Ecke 2011). This quasi-two-dimensional turbulence is a result of constraints such as rotation, density stratification, or the application of a strong magnetic field and has been the subject of many previous studies, see e.g. Chen *et al.* (2006); Alexakis & Biferale (2018); Biferale (2021); van Kan & Alexakis (2022); Alexakis (2023), and is formally defined as turbulent flow that undergoes little to no variation along a (vertical) direction of the system. This quasi-two-dimensional behaviour of turbulence can be retrieved and studied in a number of ways, depending on the type of forcing and method of analysis. The flow system central to this paper is rotating Rayleigh-Bénard convection (RRBC); an extension of the classical convection system described by Rayleigh (1916) and Bénard (1901). In RRBC, a thermal forcing is caused by a destabilising temperature difference $\Delta T \equiv T_h - T_c$ imposed between two horizontal parallel plates, which rotate around a central axis with angular velocity $\boldsymbol{\Omega}$. Here, $T_h$ and $T_c$ represent the high and low temperatures of the parallel plates, respectively. The parallel plates are spaced a distance $H$ apart and are $D$ wide. The quasi-two-dimensionality in these systems is a result of the system rotation. In these types of flows, this behaviour can be, at least partially, explained by the well-known Taylor-Proudman theorem (Proudman 1916; Taylor 1923).

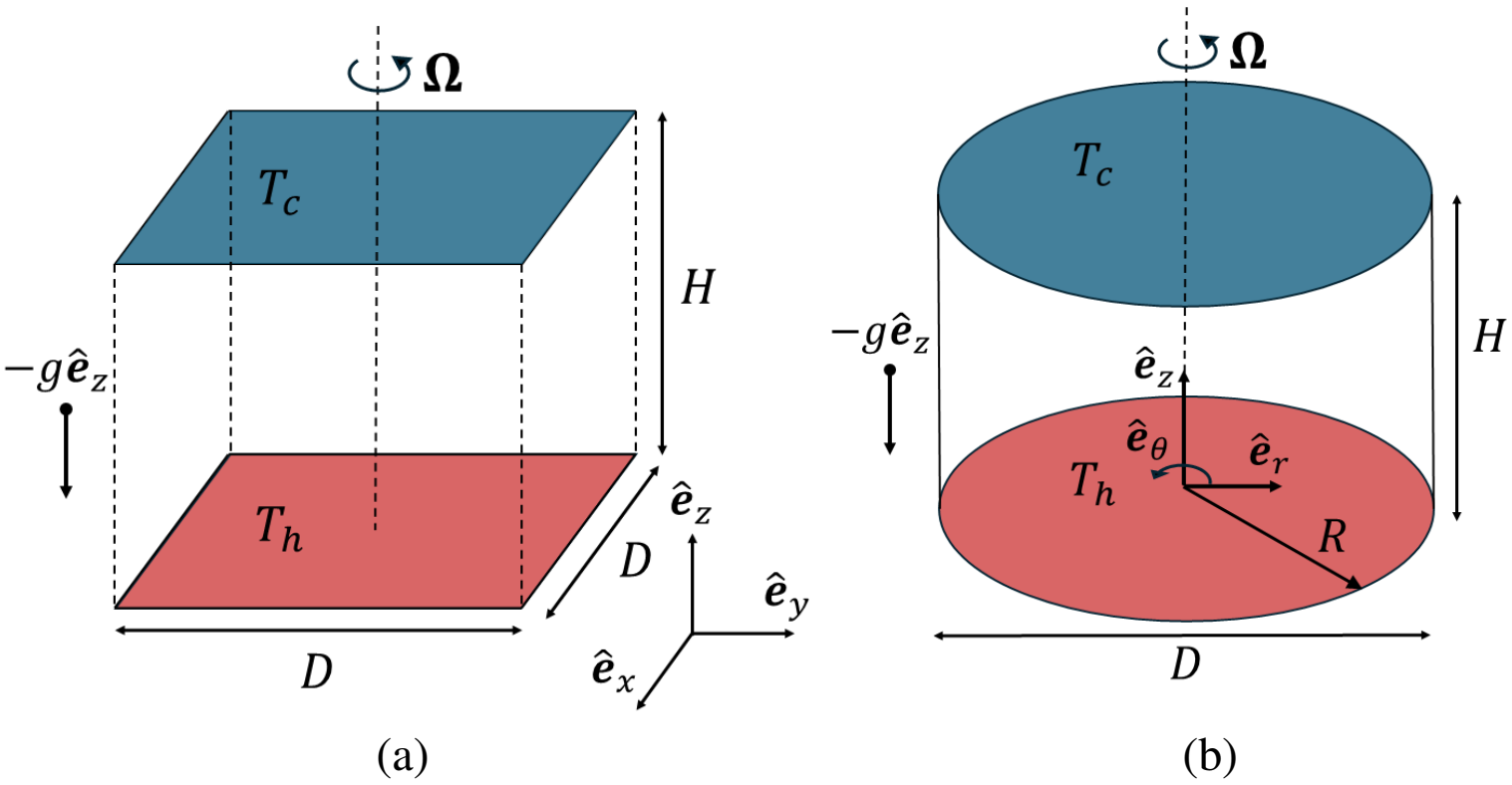

Figure 1. The domains wherein we simulate rotating Rayleigh-Bénard convection. For both the horizontally periodic domain (a) and the cylindrically bounded domain (b), this figure illustrates the geometry and the thermal forcing through a temperature difference between the top and bottom plates.

Translating this theoretical concept of energy flux to the practical calculation of this quantity in turbulent flows is not very straightforward. When we consider any flow system that does not consist of homogeneous and isotropic turbulence in a periodic box, a number of practical and pragmatic choices need to be made before even reaching the actual implementation. For RRBC, specifically, we need to consider how to handle the anisotropy and inhomogeneity of the flows itself before we can start to tackle the lack of periodicity in the domain. Having a nonperiodic domain also introduces subsequent issues that need solving, like the emergence of boundary layers. The aim of this work is to find an effective methodology for retrieving the energy flux from the velocity fields of RRBC. We consider two separate domains in which we simulate RRBC, which vary in their geometric confinement. The first of these is illustrated in Figure 1a and consists of a domain bounded from above and below by solid plates, with periodic boundary conditions imposed in the horizontal directions. The second simulation domain consists of a cylindrically bounded domain, shown in Figure 1b. The increasing complexity of the latter simulation domain serves the purpose of approaching domains of a finite nature that resemble what can be achieved in experimental research (Stevens *et al.* 2013; Kunnen 2021; Ecke & Shishkina 2023). Simulation works on horizontally periodic domains have shown that axial rotation combined with thermal forcing in RRBC can result in a sharp change between 3D and quasi-2D turbulence (Julien *et al.* 2012; Guervilly *et al.* 2014; Favier *et al.* 2014; Rubio *et al.* 2014; de Wit *et al.* 2022; de Wit 2025).

We calculate the scale-by-scale kinetic energy flux based on an associated length scale. The sign of this quantity gives us the directionality of the energy transport through that scale. When the flux is positive, energy is transported from the specified length scale towards smaller scales (corresponding to a forward or direct cascade of energy). If the calculated flux is of a negative sign, this indicates the transport of energy towards larger length scales, i.e. an inverse cascade of energy. Traditionally, the scale-by-scale flux of the kinetic energy is calculated based on Fourier decompositions of the flow (e.g. Favier *et al.* 2014). For a correct application of this, it is required that the domain has periodic boundary conditions. Since this periodicity obviously cannot be found in experimental works, another approach is required. Here, following Chen *et al.* (2006), we consider a method based on spatial filtering, inspired by its widespread application in large eddy simulation (e.g. Sagaut 2005), as an alternative to find the scale-by-scale kinetic energy flux. With this method, the Navier-Stokes equations are filtered in space with a filter kernel related to a certain spatial scale. From this, the filtered energy equation is retrieved that includes a transfer term. This transfer term gives the scale-by-scale energy flux based on the applied filtering scale. This method has been applied in this context before (e.g. Chen *et al.* 2006; Meneveau & Katz 2024; Yao *et al.* 2024).

Although these methods should give us similar information about the flow, there exists some question about the viability and generality of the results that can be obtained in practice (e.g. Yao *et al.* 2024). The analysis methods that will be treated in this paper have been individually explored and tested in earlier works, but only rarely has a comparative study been performed. One case in which a comparative study is performed is Yao *et al.* (2024), although there, the comparison is carried out in homogeneous isotropic turbulence. In our study, we compare the spatial filtering approach with a Fourier-based method as described in, e.g., Frisch (1995) in the context of rotating convection. The main purpose of our study is to perform a direct comparison of methods to calculate the energy flux specifically within the inhomogeneous and anisotropic realm of (partially) confined RRBC.

### 1.1. *Governing parameters and non-dimensionalization*

The dimensionless numbers that govern RRBC are: The Rayleigh number, the Ekman number and the Prandtl number. These are defined as:

$$Ra \equiv g\alpha \Delta T H^3/\nu\kappa, \tag{1.1}$$

$$Ek \equiv \nu/2\Omega H^2, \tag{1.2}$$

$$Pr \equiv \nu/\kappa. \tag{1.3}$$

In these definitions $\nu$, $\kappa$, $g$, and $\alpha$ respectively represent the kinematic viscosity, thermal diffusivity, gravitational acceleration, and thermal expansion coefficient. The Rayleigh number is the ratio between the buoyancy and the dissipation, the Ekman number relates the viscous forces to the Coriolis forces, and the Prandtl number relates the viscous forces to the thermal dissipation. Finally, we define the aspect ratio of the system as the fraction of the domain width over the domain height: $\Gamma \equiv D/H$. The Prandtl number is set to unity in all simulations. With the definition of the dimensional parameters that control the solutions of the governing equations and the dimensionless numbers complete, we turn our attention to Figure 2. This figure, which is based on Favier *et al.* (2014), denotes the parameter space of the Rayleigh and Ekman numbers for horizontally periodic RRBC in which we expect to find a so-called Large-Scale Vortex (LSV) that is resulting from the inverse cascade. In this figure, the simulations of Favier *et al.* (2014), Guervilly *et al.* (2014) and Kunnen *et al.* (2016) are also shown. Since the overarching goal of this work is to be able to quantify the scale-by-scale flux of kinetic energy in RRBC, we focus on systems which have at least a partly inverse energy flux, as this makes for an energy flux that is both positive and negative. For reference, we also show the cylindrical simulations that are the subject of Section 4 in Figure 2. Note that this regime diagram only applies to horizontally periodic domains, the effect of the added confinement in cylindrically bounded domains will significantly alter the flow behaviour.

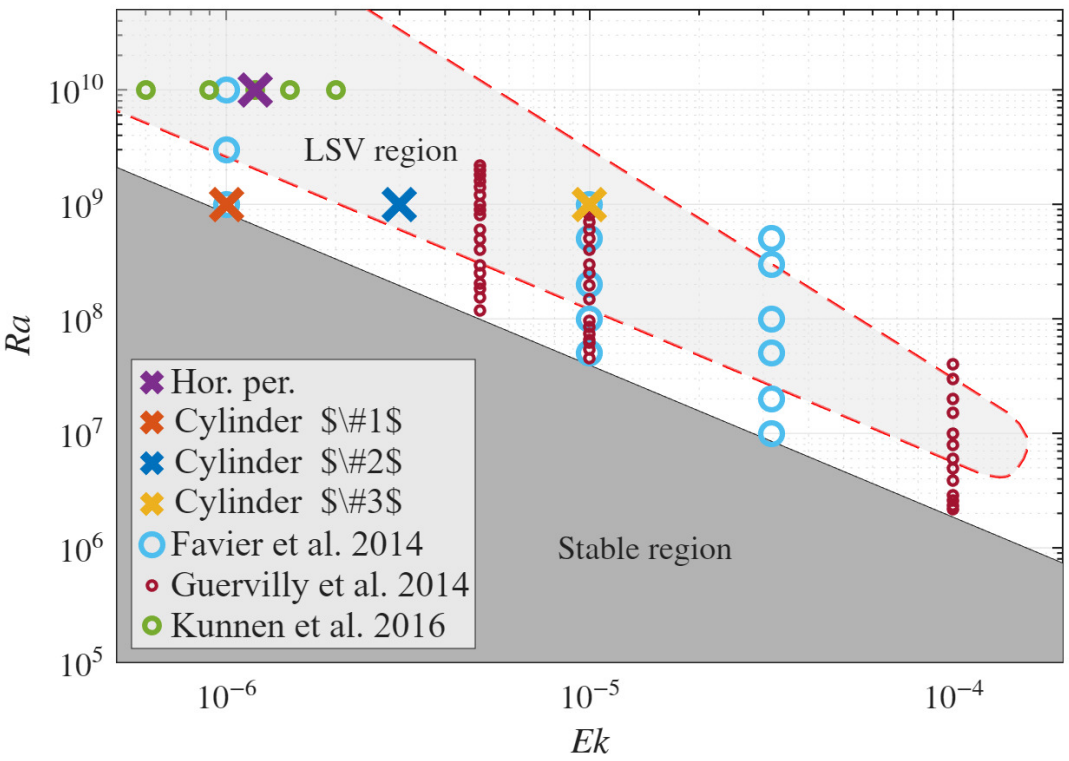

Figure 2. The parameter regime of Rayleigh and Ekman numbers where we expect large-scale vortices to form in horizontally periodic simulations of RRBC. Based on figure 7 from Favier *et al.* (2014). The dark shaded region in the bottom right corner denotes the stable region where we expect to see no convection. We denote the simulations performed in this project by coloured crosses and the cases treated in Favier *et al.* (2014); Guervilly *et al.* (2014); Kunnen *et al.* (2016), using coloured circles.

## 2. Theory and methodology

To describe rotating Rayleigh-Bénard convection, the Navier-Stokes equations are supplemented by the heat equation and the statement of incompressibility. This is combined with the assumptions of a linear dependence of the density on the temperature, with all other fluid properties considered constant and independent of temperature (Chandrasekhar 1961). With this, the governing equations are presented within the Oberbeck-Boussinesq approximation as

$$\frac{\partial \boldsymbol{u}}{\partial t} + \boldsymbol{u} \cdot \boldsymbol{\nabla} \boldsymbol{u} = -\boldsymbol{\nabla} p + \sqrt{\frac{Pr}{Ra}} \nabla^2 \boldsymbol{u} + T \hat{\boldsymbol{e}}_z - \frac{1}{Ek} \sqrt{\frac{Pr}{Ra}} \hat{\boldsymbol{e}}_z \times \boldsymbol{u}, \tag{2.1a}$$

$$\frac{\partial T}{\partial t} + \boldsymbol{u} \cdot \boldsymbol{\nabla} T = \frac{1}{\sqrt{PrRa}} \nabla^2 T, \tag{2.1b}$$

$$\boldsymbol{\nabla} \cdot \boldsymbol{u} = 0. \tag{2.1c}$$

All parameters have been non-dimensionalized using length scale $H$, convective velocity scale $U \equiv \sqrt{g\alpha \Delta T H}$ and temperature scale $\Delta T$. All results in this paper are presented in dimensionless form using these scales for nondimensionalisation, i.e. the dimensional wave number $\tilde{k}$ and time $\tilde{t}$ are defined as $\tilde{k} \equiv k/H$ and $\tilde{t} \equiv tH/U$ using their dimensionless counterparts $k$ and $t$.

We will be using Direct Numerical Simulation (DNS) to study RRBC. Both codes used in this work use second-order accurate finite-difference discretisation and third-order accurate Runge-Kutta timestepping. The time step is determined dynamically based on the Courant-Friedrichs-Lewy (CFL) condition, a maximum CFL of 1.2 is used as an upper limit. With this condition, which falls well within the theoretical CFL limit of $\sqrt{3}$ for a third-order Runge-Kutta timestepping scheme (Spalart *et al.* 1991), the time step is of $O(1 \times 10^{-3})$. The simulation of the horizontally periodic domain is placed on a Cartesian grid, while the cylindrically bounded simulations are placed on a cylindrical coordinate system. The code for the horizontally periodic domain is based on an earlier implementation for cylindrical domains (Verzicco & Orlandi 1996; Verzicco & Camussi 2003); further details can be found in Ostilla-Monico *et al.* (2015). We apply a stress-free boundary condition on all nonperiodic bounding surfaces to promote the upscale energy transfer and large-scale flow formation. This is based on the comparison of no-slip and stress-free boundary conditions on the bounding plates in works such as Kunnen *et al.* (2016); Plumley *et al.* (2016); Aguirre Guzmán *et al.* (2020); Song *et al.* (2024*a*). This is supplemented by a nonpermeability condition at the bounding surfaces. To ensure that the smallest active scales are resolved, the grid and resolution are specifically chosen to resolve the Kolmogorov and Batchelor length scales for the velocity and temperature fields (Kolmogorov 1941; Batchelor 1959). For the boundary regions, where the flow displays the strongest spatial variations, the vertical grid points are spaced non-uniformly using a hyperbolic tangent based function,

$$z(c) \equiv \frac{1}{2} + \frac{\tanh\left(\frac{2c - n_z - 2}{n_z}\beta\right)}{2\tanh(\beta)}, \tag{2.2}$$

where $z$ and $c$ represent the vertical location and cell index (with $c$ ranging from 1 to $n_z$, the number of grid cells in the vertical direction), respectively. $\beta$ is used as a tuning parameter. For our simulations a tuning parameter of $\beta = 4$ was used to achieve a locally higher resolution near the top and bottom boundaries.

### 2.1. *The scale-by-scale energy flux from the Fourier-based method*

For the conception of the first analysis method, the derivation presented in the book by Frisch (1995) is followed. This method relies heavily on the usage of Fourier transformations and the characteristics associated with these, which is why we shall refer to this as the *Fourier-based* method. With this method, the velocity fields are split into two parts, based on a chosen wavenumber $k_0$, as

$$\begin{aligned} f^{<}_{k_0}(\boldsymbol{x}) &\equiv \sum_{|\boldsymbol{k}|\leq k_0} \hat{f}_{\boldsymbol{k}} e^{i\boldsymbol{k}\cdot\boldsymbol{x}}, \\ f^{>}_{k_0}(\boldsymbol{x}) &\equiv \sum_{|\boldsymbol{k}|> k_0} \hat{f}_{\boldsymbol{k}} e^{i\boldsymbol{k}\cdot\boldsymbol{x}}. \end{aligned} \tag{2.3}$$

Here we denote the Fourier transformed field with the use of the circumflex accent. In Equation (2.3) we have defined two filtered fields: i. $f^{<}_{k_0}(\boldsymbol{x})$, which is the field with all wavenumbers $\boldsymbol{k}$ with length up to and including $k_0$, and ii. $f^{>}_{k_0}(\boldsymbol{x})$, which is the field with all wavenumbers greater than $k_0$. These definitions have the consequence that the two filtered fields add up to the original: $f(\boldsymbol{x}) = f^{<}_{k_0}(\boldsymbol{x}) + f^{>}_{k_0}(\boldsymbol{x})$. Applying this to the Navier-Stokes equations and taking the dot product with the low-pass filtered velocity field, we obtain the kinetic energy equation. From this kinetic energy equation we identify the kinetic energy flux across scales with wavenumber $k_0$ as

$$\Pi_{k_0} \equiv \left\langle \boldsymbol{u}^{<}_{k_0} \cdot (\boldsymbol{u}\cdot\boldsymbol{\nabla})\, \boldsymbol{u}^{>}_{k_0} \right\rangle, \tag{2.4}$$

where the angled brackets denote a spatial average over the domain taken to find a single value of $\Pi_{k_0}$. This formulation is a rewrite of the result that can be found in equation (2.52) of Frisch (1995). A positive sign for $\Pi_{k_0}$ indicates a transfer of energy to the smaller scales (larger wavenumber) and a negative sign indicates an inverse flux of kinetic energy toward the larger scales (smaller wavenumber). The recipe is thus as follows: a low-pass filtering of a given velocity field is performed, after which, the low-pass filtered, the 'remainder', and the original velocity fields can be used in Equation (2.4) to find a value for $\Pi_{k_0}$.

In the application of this method to the three-dimensional velocity fields, we perform the Fourier transformation of the field in 2D on horizontal slices of the domain. These are the homogeneous and periodic directions of the computational domain. For the cylindrically bounded simulations of Section 4, some changes are made to mimic a similar level of homogeneity and periodicity in the horizontal directions of the computational domain. The specific changes made are detailed in the introduction of Section 4. The Fourier modes that are selected consist of the grid points in Fourier space that are chosen based on the length of the wavenumber vector $\boldsymbol{k}$. Since we perform the initial step of the Fourier filtering in 2D, this process of selecting shells of a specific wavenumber interval $k_0 \leq |\boldsymbol{k}| < k_0 + \Delta k$ (as shown in Section 2.1 of Yao *et al.* 2024) is performed on a 2D plane. We have validated that the definition of the wavenumber interval does not significantly affect the outcome of the analysis. The resulting slices are subsequently combined into a 3D array and used in the subsequent steps of the analysis. For a more detailed review of the effect of the two-dimensionalisation of this process, we refer to Section 4.2.1 of Lemm (2024). With this adaption, the gradient in Equation (2.4) consists of spatial derivatives calculated in Fourier space for the horizontal coordinate directions, and we use a $4^{\text{th}}$ order accurate central differencing scheme for the finite difference spatial derivatives in the vertical coordinate direction.

### 2.2. *The scale-by-scale energy flux from the spatial filtering method*

The second analysis method is based on the concept of spatial filtering. This filtering takes place entirely in real space and is most commonly used in large eddy simulation (LES, Sagaut 2005). This simulation method reduces the computational requirements of numerically solving the Navier-Stokes equations in full (as is done in DNS) by only focussing on the behaviour of the large length scales and replacing the small-scale dynamics with a so-called subgrid model. The small-scale behaviours are taken out using a spatial filter. Such a filter can be imagined as an operator that splits the velocity fields into two: one part is retained representing the larger-scale dynamics and one part representing dynamics below the so-called filter scale. We follow the books by Sagaut (2005) and Pope (2000) for the derivation of this analysis method. With the spatial filtering method, the velocity fields are filtered using a convolution operation:

$$\overline{\boldsymbol{u}}(\boldsymbol{x}) \equiv \iint G(\boldsymbol{r})\boldsymbol{u}(\boldsymbol{x}-\boldsymbol{r})\mathrm{d}\boldsymbol{r}. \tag{2.5}$$

This operation uses a filter function denoted by $G(\boldsymbol{r})$. The filter function $G(\boldsymbol{r})$, is a 2D function, since we perform this filtering operation in the two homogeneous directions, i.e. on two-dimensional slices of the 3D velocity field. Throughout this paper, we will only consider a Gaussian-shaped filter,

$$G_{\mathrm{Gauss}}(\boldsymbol{x}) \equiv \frac{6}{\pi\Delta^2}\exp\left(\frac{-6|\boldsymbol{x}|^2}{\Delta^2}\right). \tag{2.6}$$

The filter width $\Delta$ is used to define the spatial extent of the filtering operation. The prefactor in Equation (2.6) is chosen to achieve second moment equivalence of the Gaussian filter and a box-shaped filter of width $\Delta$ (Pope 2000). The analysis of the effects of different filter shape implementations, such as a box-shaped or a sharp-spectral filter, can be found in Section 4.2.2 of Lemm (2024) and in Section B where we show a comparison of the spatial filtering kinetic energy flux calculated using a Gaussian and a box-shaped filter. These results illustrate that the results of the spatial filtering method do not depend significantly on the choice between a Gaussian and a box-shaped filter.

With spatial filtering, the velocity field is split into two parts: the filtered velocity field and the residual field. Together, these return the original velocity field as

$$\boldsymbol{u}(\boldsymbol{x},t) = \overline{\boldsymbol{u}}(\boldsymbol{x},t) + \boldsymbol{u}'(\boldsymbol{x},t), \tag{2.7}$$

where $\boldsymbol{u}'(\boldsymbol{x},t)$ is the residual velocity field. In contrast to the Fourier-based method, this residual field does not play a role in the determination of the energy flux with the spatial filtering method. With the velocity field promptly filtered, we follow the derivation of the energy flux found in chapters 5 and 6 of Sagaut (2005) or in chapter 13.3 of Pope (2000). Following the derivations in these works, we can identify the scale-by-scale kinetic energy flux as

$$\Pi_r \equiv \left\langle -\boldsymbol{\tau}^r : \overline{\boldsymbol{S}} \right\rangle, \tag{2.8}$$

where the double dot product denotes a Frobenius inner product, and the angled brackets again denote the spatial average over the simulation domain. In this definition, we identify the residual stress tensor and the filtered rate of strain tensor as

$$\boldsymbol{\tau}^r \equiv \overline{\boldsymbol{u}\boldsymbol{u}} - \overline{\boldsymbol{u}}\ \overline{\boldsymbol{u}}, \tag{2.9}$$

$$\overline{\boldsymbol{S}} \equiv \frac{1}{2}\left(\boldsymbol{\nabla}\overline{\boldsymbol{u}} - (\boldsymbol{\nabla}\overline{\boldsymbol{u}})^T\right). \tag{2.10}$$

The residual stress tensor is similar to the Reynolds stress tensor except that it is based on the filtered velocity field, while the Reynolds stress tensor is based on the time-averaged field (Pope 2000). In this approach, the flux term represents the transport of kinetic energy between the filtered and residual velocity fields, again with the angled bracket indicating the average over the domain. The length scale at which the energy flux is calculated is the filter width $\Delta$.

### 2.3. *Comparison*

If we put these two known methods for calculating the energy flux side-by-side, we can *a priori* identify two main differences. First, we know that both analysis methods use some form of a filtered velocity field to compute the flux as Equations (2.4) and (2.8), respectively. In Section A, we show that these definitions become analytically equivalent when spatially averaged, but only if the applied filter can be considered a Reynolds operator. While the sharp spectral filter of the Fourier-based method is a Reynolds operator, the Gaussian filter not strictly, so that for the latter, only Equation (2.8) is valid. The second difference results from the difference in the filtering operation itself, which becomes evident when comparing the transfer functions in spectral space. Fourier filtering is sharp in spectral space:

$$\hat{G}_{\text{Fourier}}(\boldsymbol{k}) \equiv \mathcal{H}(k_c - |\boldsymbol{k}|), \tag{2.11}$$

where $\mathcal{H}$ denotes the Heaviside step function and $k_c$ the cut-off wave number. This is functionally equivalent to Equation (2.3). From this we see that the Gaussian filter is effectively a smoothed version with transfer function

$$\hat{G}_{\text{Gauss}}(\boldsymbol{k}) \equiv \exp\left(-\frac{|\boldsymbol{k}|^2\Delta^2}{24}\right), \tag{2.12}$$

decaying smoothly from 1 to 0 around the wavenumber $|\boldsymbol{k}| = \pi/\Delta$. To use a filter width $\Delta$ in the spatial filtering method that matches the wave number from the Fourier-based method, we will thus use this relation to compare the results of the analysis methods:

$$k_c \equiv \pi/\Delta. \tag{2.13}$$

We can thus anticipate comparable results between both both methods, with differences resulting from either the difference in the filtering operation and the numerical implementations of the methods. Finally, we point out that in the application of the spatial filtering method, the filter width $\Delta$ can be chosen to be of any arbitrary size. With this, the energy flux can be calculated at any spatial scale. This is in contrast to the Fourier-based method, for which the choice of wavenumbers is restricted to integer multiples of the smallest wavenumber $k_{min}$. The application of both analysis methods in the upcoming sections will be used to find out to what extent the similarities between the methods will hold up and where differences emerge.

## 3. Rotating Rayleigh-Bénard convection in a horizontally periodic domain

In the previous section, we introduced the methodology to simulate RRBC together with the tools to analyse the velocity fields resulting from these simulations. The first simulation domain is periodic in the horizontal coordinate directions. This section consists of a brief introduction and a validation of the performed simulation, followed by a detailed discussion of the scale-by-scale kinetic energy flux found using the two methods of analysis. As mentioned above, our objective was to obtain a simulation that incorporates a significant inverse energy flux. Previous works such as Kunnen *et al.* (2016); Favier *et al.* (2014) were consulted to select a set of system parameters that are known to lead to the formation of an LSV. The Rayleigh number is set to $Ra = 10^{10}$ and the Ekman number to $Ek = 1.2\times10^{-6}$. This range of parameter values has seen similar structures form in van Kan *et al.* (2025). The aspect ratio of the domain is set to $\Gamma = 0.51$. With this, the smallest (dimensionless) wavenumber to be found with the Fourier-based method becomes: $k_{min} \equiv 2\pi/\Gamma = 12.32$. This is a direct result of the methodology described in Section 2.1. The resolution is set to $(n_x, n_y, n_z) = (512, 512, 960)$ and with that, the minimum requirement for the resolution of the smallest active scales is met. With this resolution a minimum and maximum grid spacing of $1.0\eta_k$ and $2\eta_k$ at the bounding plates and domain middle is achieved, where $\eta_k$ represent the smallest active scale or the Kolmogorov length scale. In the horizontal direction the chosen resolution results in a grid spacing of $0.9\eta_k$.

### 3.1. *Flow characterisation and stationarity*

In our analysis, we ensure that statistics are collected in the statistically stationary state. An *a posteriori* parameter that was monitored to gauge this stationarity is the Nusselt number, which is defined as the ratio between the convective and the conductive heat transport in the system: $Nu \equiv qH/k_T\Delta T$, where $q$ and $k_T$ represent the averaged total heat flux and the thermal conductivity of the fluid. A running average over the simulation run results in $Nu \approx 74$, as can be seen in Figure 3a. This is within proximity of the result of $Nu = 68.5$ from the simulation performed by Kunnen *et al.* (2016). The difference between the found Nusselt numbers in this paper and in the simulation we performed most likely results from the almost doubling of the resolution and due to the significantly longer run time employed here; the LSV saturates quite slowly. A different metric that gives insight into the stationarity of the simulation are the root mean square (RMS) values of the velocity field. In Figure 3b we show the RMS velocities for each of the Cartesian velocity components. From this figure, we see that after $t \approx 500$ the domain-averaged RMS velocities become statistically stationary.

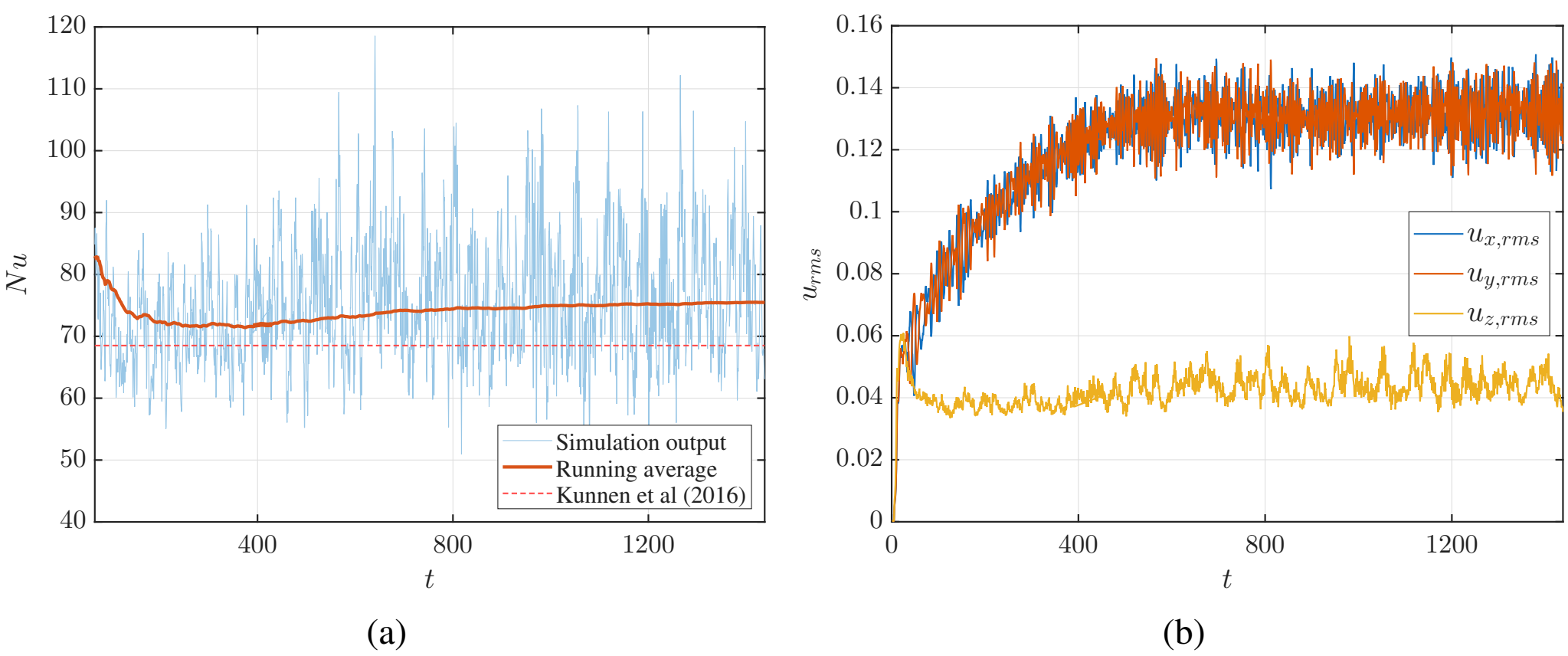

Figure 3. (a) The Nusselt number output in blue with the running average of this output in orange and the Nusselt number from Kunnen *et al.* (2016) with the red dashed line. (b) The root mean square velocities of the horizontally periodic simulation. The horizontal components of the velocity ($u_{x,rms}$, $u_{y,rms}$) reach a steady state after about $t = 500$, the Nusselt number increases slightly until $t \approx 1000$.

Another metric we can extract from the velocity data is the kinetic energy spectrum shown in Figure 4. This figure shows the averaged energy spectrum in the domain at the end of the simulation run. The energy spectrum is shown in three varieties: energy $\hat{E}_{hor} \equiv \frac{1}{2}(\hat{u}_x^2 + \hat{u}_y^2)$ in horizontal flow; energy $\hat{E}_{vert} \equiv \frac{1}{2}\hat{u}_z^2$ in vertical flow; and total energy $\hat{E}_{tot} \equiv \hat{E}_{hor} + \hat{E}_{vert}$.

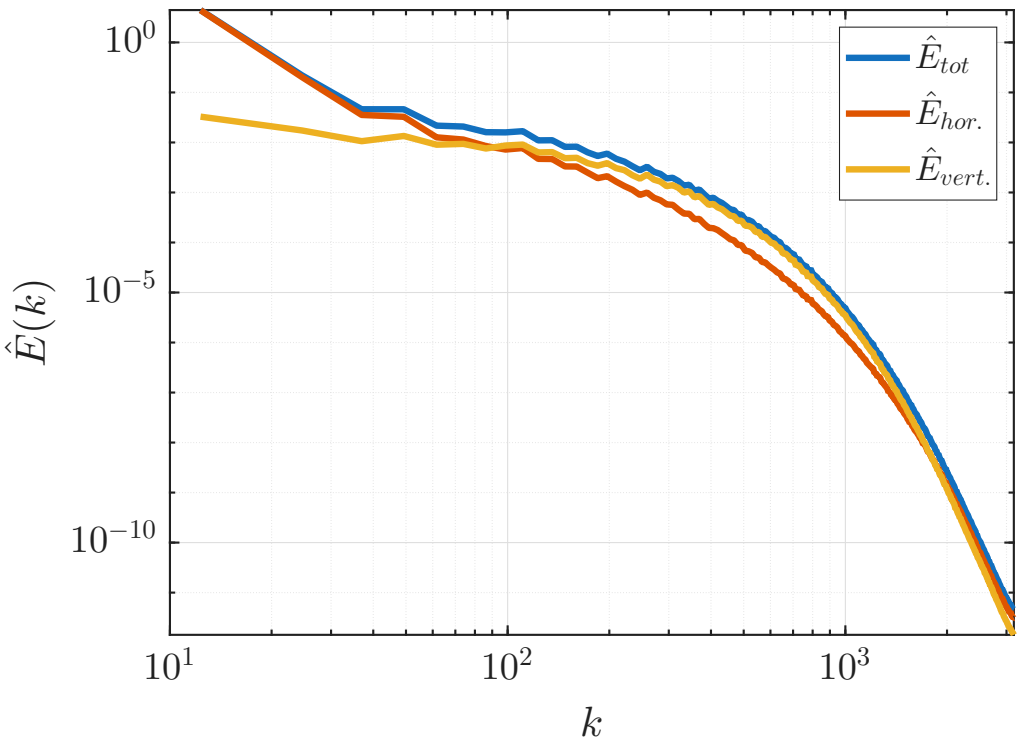

Figure 4. The kinetic energy spectra shown for a single instantaneous velocity field of the horizontally periodic RRBC simulation. The kinetic energy is shown as the total ($\hat{E}_{tot}$), as well as the horizontal ($\hat{E}_{hor.}$) and vertical ($\hat{E}_{vert.}$) components.

Considering the total kinetic energy, we see that there is a clear peak at the smallest wavenumbers or at the largest length scales in the system. This indicates that the flow is dominated by structures that encompass the majority of the simulation domain. We also note that at the smaller wavenumber values, the split between the horizontal and vertical kinetic energy curves is overwhelmingly in favour of the horizontal component. This shows not only that most energy is concentrated at the largest length scales but also that this is predominantly in the horizontal plane, which already hints at the presence of an LSV. On the basis of the analysis of the energy spectrum, one would expect to see flow structures that are orientated in the horizontal direction and that cover a large portion of the horizontal simulation domain. A way of visualising the flow structure using the vorticity is shown in Figure 5. We show the vertical component of the vorticity: $\omega_z$, of four horizontal slices of the domain at different heights. The vertical vorticity confirms our predictions based on the energy spectra and the RMS velocities: the presence of a large-scale domain-filling vortex.

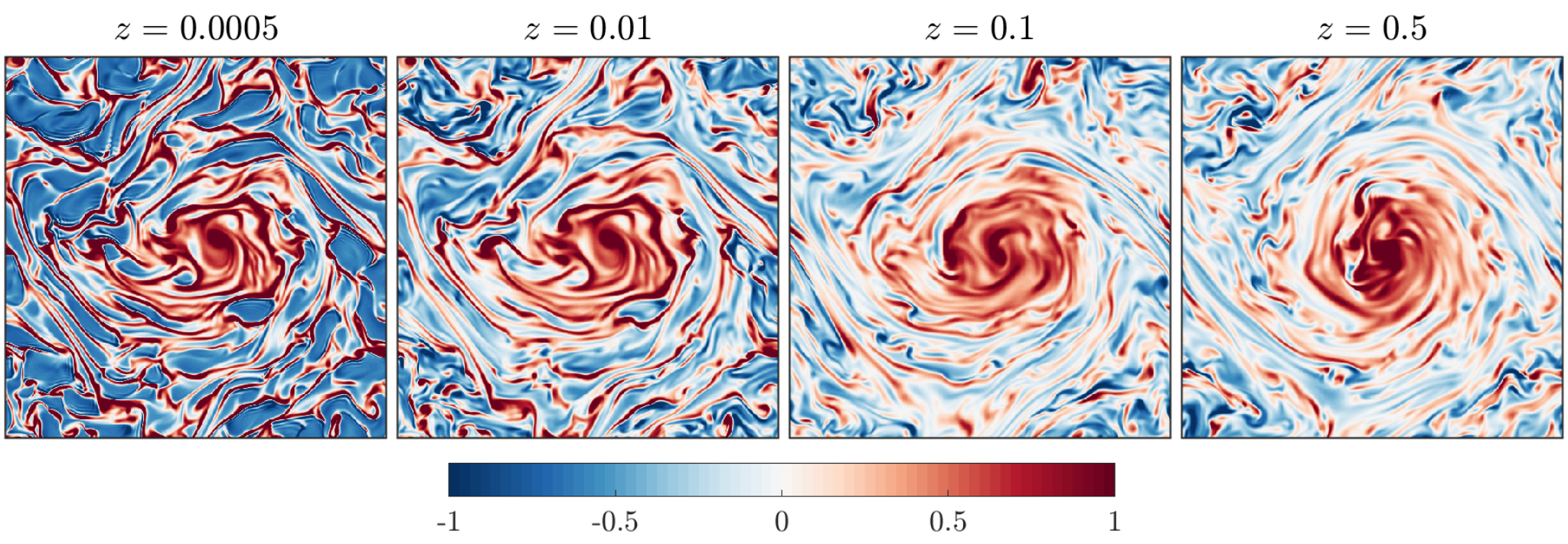

Figure 5. The vertical vorticity $\omega_z$ shown for four horizontal slices of the lower half of the simulation domain. The vertical vorticity of the flow shows a lot of similarities between these vertical coordinate locations, some characteristics of a quasi-2D flow, like the size and shape of the LSV, do vary over the vertical coordinate locations.

From this figure, we can observe a number of things. Starting by looking at the leftmost panel, very close to the bottom plate, we observe that the LSV is still clearly present here but also that there are smaller-scale plume-like structures, as also seen in previous work by Aguirre Guzmán *et al.* (2020, 2021, 2022); Song *et al.* (2024*b*).

### 3.2. *The scale-by-scale kinetic energy flux*

To quantitatively compare the proposed analysis methods, spatial and temporal averaging are employed. The analysis methods are applied to a series of 25 instantaneous velocity fields which are taken with a constant interval of $t = 5$. This is done to make sure that the fields are sufficiently decorrelated in time. The sensitivity to velocity fluctuations can be observed by looking at the domain-averaged flux curves of the individual instantaneous velocity fields in Figure 6. The individual flux curves (light shaded and solid lines) together form the series which results in the temporally averaged scale-by-scale energy flux curves, which are indicated in this figure by thick dashed lines. In the remainder of this paper we will consistently denote the energy flux resulting from the spatial filtering method in blue, the energy flux resulting from the Fourier-based method in orange, and the cross-over between the forward and inverse cascades at $\Pi = 0$ with a red dashed line. In Figure 6 we observe that both methods agree largely on the resulting energy flux, which can be summarised as: $\Pi(k \lesssim 37) < 0$ and $\Pi(k \gtrsim 37) > 0$, indicating a change in the energy cascade from inverse to forward at $k \approx 37$.

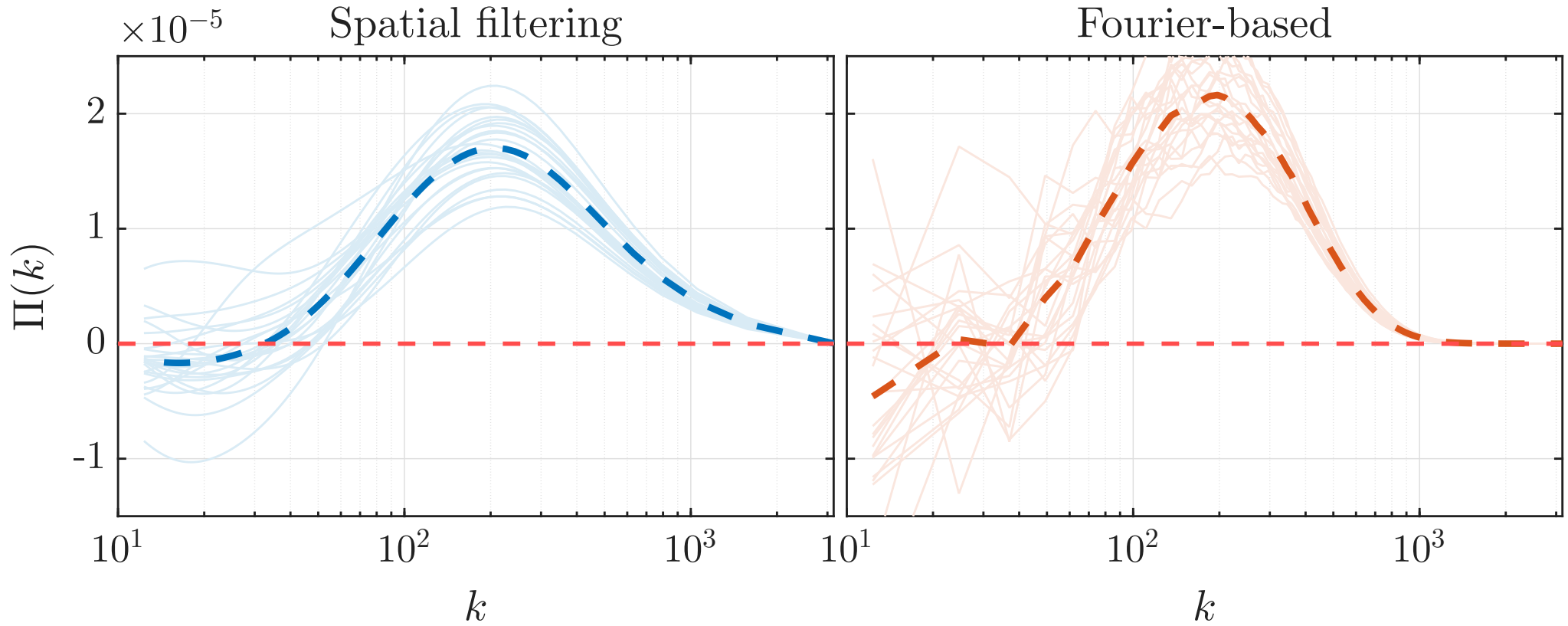

Figure 6. The domain averaged scale-by-scale kinetic energy flux curves of each of the individual instantaneous velocity fields. The spatial filtering method in the left panel, the Fourier-based method in the right panel. The light-shaded lines show the individual results from each of the velocity fields, the blue and orange dashed lines indicate the temporal averages over the series of velocity fields.

Looking at the individual energy flux curves in Figure 6, we can see that some of these have a positive energy flux for the entire wavenumber range. This indicates that the accumulation of energy at larger length scales is irregularly interrupted by moments in which the direct energy cascade takes over. These interruptions in the inverse energy cascade signify moments in which the transport of energy towards smaller scales takes the upper hand and counteracts the growth of the LSV.

The Fourier-based method results in a flux curve that is more erratic and has a larger overall spread, whereas the spatial filtering results seem more smooth with less variation overall. The direct comparison between the two analysis methods is shown in Figure 7. To give an indication of the spread of the energy flux results in the time series, a light-shaded area around the flux curves is added to indicate the 95% confidence interval of the series of energy flux curves that form the temporal average. This figure can be divided into two parts: the first part below the red dashed line where $\Pi < 0$, and the second part that starts after the dashed line is crossed where the energy flux becomes positive. We see that both methods agree that the inverse part of the cascade spans length scales that correspond roughly to $k \leq 37$. This translates to wavelengths that span at least a third of the horizontal size of the domain. For wavenumbers larger than this, both methods show a positive value for the energy flux, indicating that at these wavenumbers, the net transport of kinetic energy is towards smaller scales.

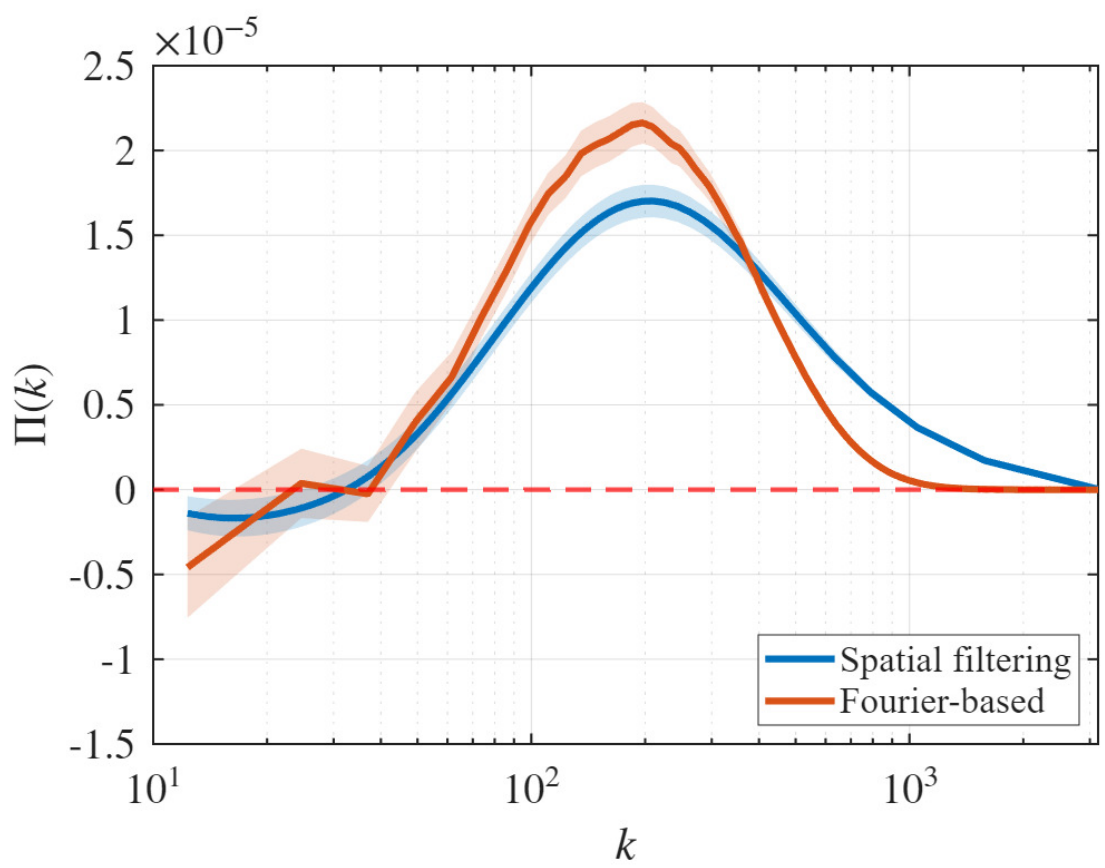

Figure 7. The spatial- and temporal averaged scale-by-scale flux of kinetic energy of both the spatial filtering and the Fourier-based method. The convergence between both methods is made visible here, combined with the shaded area around the energy flux curves that represents the 95% confidence interval.

With these results, we find a compelling agreement between the two analysis methods. Although the flux curves are not identical, they do convey largely the same qualitative information. When we look more in-depth at the differences between the distinct analysis methods, two things stand out. The first observation is that the largest difference in the flux amplitude is found in the forward cascade region. The second observation lies with the aforementioned inverse energy cascade region. Given the overlap of the 95% confidence intervals in this region, we conclude that there is no significant disagreement between the two approaches here. However, we do see a significantly broader spread for the Fourier-based method compared to the spatial filtering method. This difference can be explained by the implementation of the analysis methods. The Fourier-based method relies on the Fourier ring filtering described in Section 2.1 of Yao *et al.* (2024) (as well as in Section 3.3.1 of Lemm (2024)). The wavenumbers are grouped based on 2D binning into circular shells in Fourier space. The bin edges are integer multiples of $k_{min}$. Because of this, only very few data points in spectral space contribute to the low wavenumber shells which is likely to be the cause of larger spread we see in the results. Additionally, the $\boldsymbol{k}$ vectors that contribute to these low-$k$ shells can be of rather different length (for example, the vectors $(1, 1)k_{min}$ and $(2, 0)k_{min}$ contribute to the same shell but differ by a factor $\sqrt{2}$ in length). This is in contrast to the spatial filtering method. With this method the choice of filter width $\varDelta$ is completely arbitrary, which allows any wavenumber value to be analysed without this being related to the resolution or grid.

### 3.2.1. *Vertical distribution of kinetic energy flux*

Until this point, we have only considered the scale-by-scale energy flux as a quantity averaged over the entire domain. Because we know that the turbulence found in RRBC is of an inhomogeneous nature, it can be insightful to look at the way the energy flux varies along the vertical coordinate of the system; this is shown in Figure 8 for $\Pi(k = k_{min} = 12.32)$. Here we plot the horizontally averaged energy flux against the vertical coordinate of the domain.

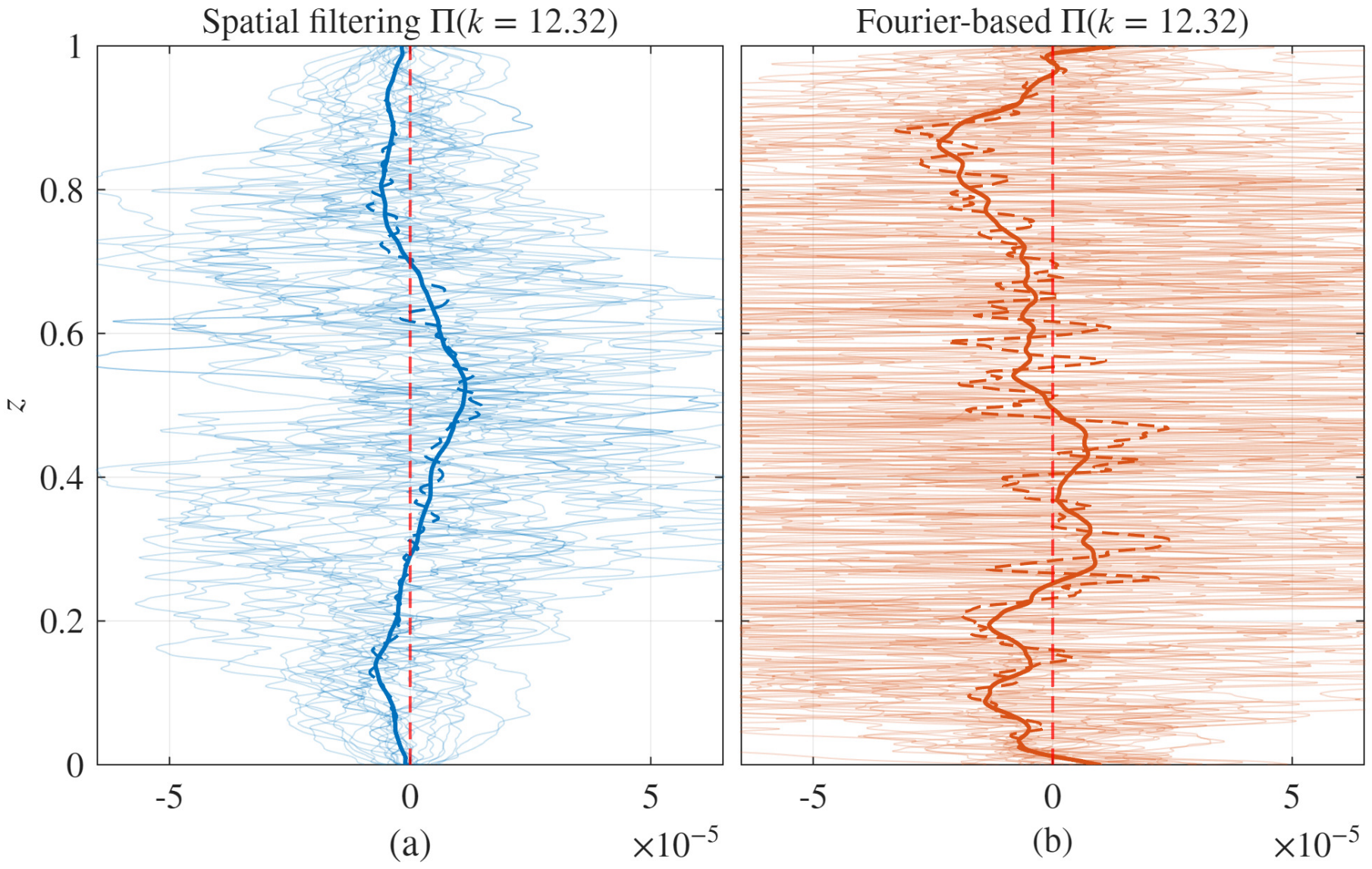

Figure 8. The spread of the horizontally averaged energy flux curves over the time series shown in the light shaded curves, the temporal average over the entire time series as the dashed line and the moving average over the $z$-coordinate with the solid line. The spatial filtering (a) and the Fourier-based method (b) both result in a similar distribution over the vertical coordinate, while the spread of the instantaneous flux curves varies significantly for this case of $k = k_{min} = 12.32$.

As was the case with the domain-averaged energy flux, we observe significant fluctuation between the individual points in time when the energy flux is plotted against the vertical coordinate. Each of the panels contains the entire spread of the time series of energy flux curves, as well as their temporal average shown with the dark dashed lines. Finally, we also show a moving average over the z-coordinate with solid lines. This moving average gives a clearer indication of the overall trend of the energy flux throughout the vertical extent of the system. From this figure, we can see that the overall spread in the time series is larger for the Fourier-based method. This is also seen in the temporal average curve, which is more dominated by large peaks and valleys compared to the spatial filtering temporal average. A similar procedure was followed for $k = k_{min} \times [1, 2, 4, 8] = [12.32, 24.64, 49.28, 98.56]$ in Figure 9, although here we only show the moving averages over the z-coordinates combined with the 95% confidence intervals over the time series. The convergence of the two analysis methods is emphasised for higher wavenumber values, with the best match seen in this figure for $k = 98.56$.

Looking at the vertical distribution of the scale-by-scale kinetic energy flux for wavenumbers in the inverse cascade range of $k \leq 37$ (and to a lesser extent for the higher wavenumber values), we observe a tendency of the energy flux to be negative mainly in the regions of the simulation domain near the top and bottom plates. We see that for $k = 12.32$ and $k = 24.64$ the flux is negative in the top and bottom twenty to twenty-five percent of the domain, while it remains largely positive in the middle region of the domain. We anticipate that mergers of the vortical Ekman plumes generated near the bottom and top plates (e.g. Julien *et al.* 1996, 2012) are taking place at these locations; a physical mechanism that can explain the occurrence of inverse energy transfer there. Within the thermal boundary layer, the energy flux is clearly positive for the results from the Fourier-based method but not for the spatial filtering method. Why this occurs remains unexplained and warrants further research.

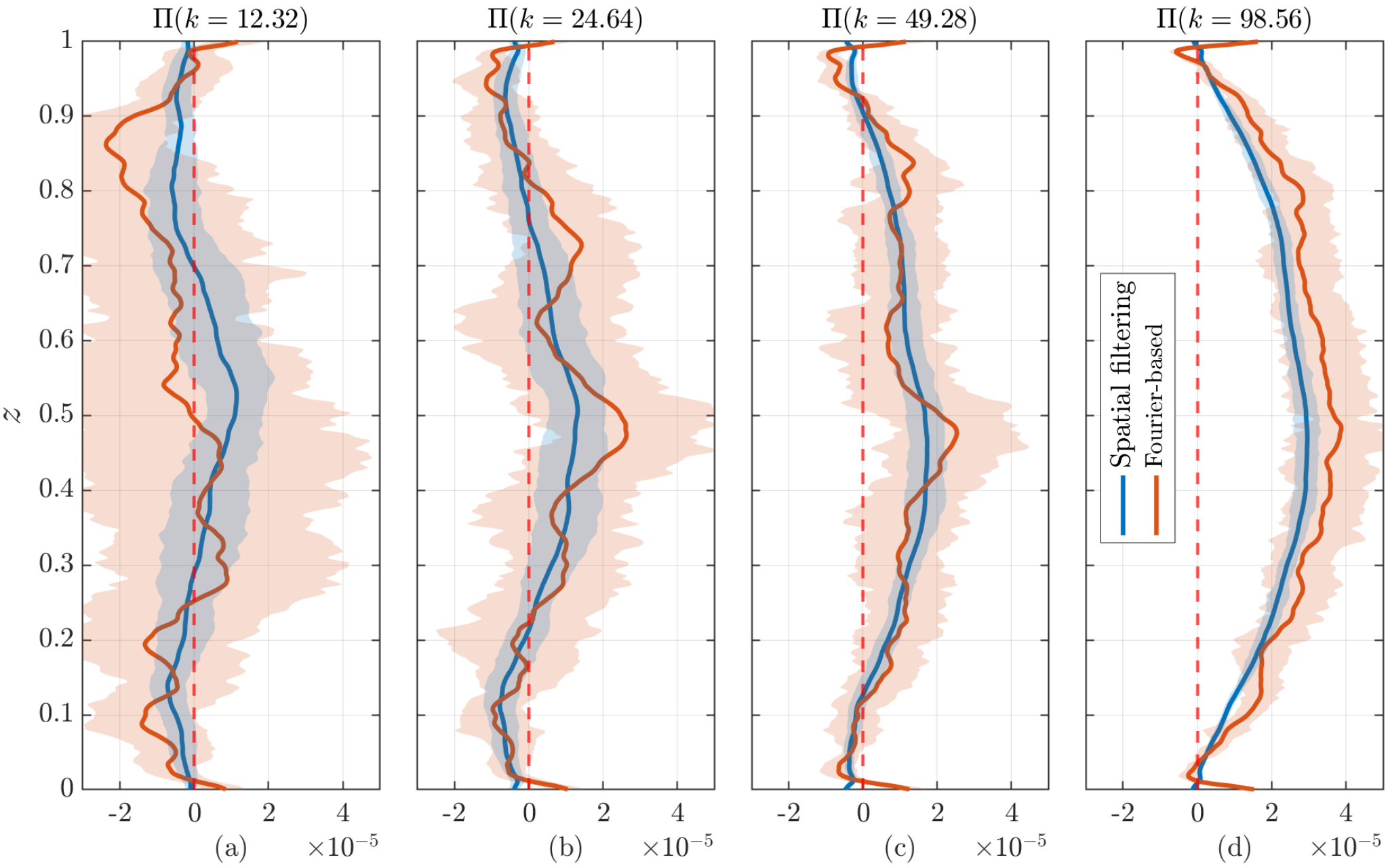

Figure 9. The vertical distribution of the energy flux resulting from both analysis methods is shown for four different wavenumbers, together with the 95% confidence interval over the time series.

### 3.2.2. *Horizontal distribution of kinetic energy flux*

To take a closer look at how the resulting fluxes from both methods differ, we examine the raw data without any averaging. Since the kinetic energy flux is a local definition in real space that is calculated at each of the grid points individually, we can visually compare the horizontal distribution of the energy flux resulting from both methods in slices of the simulation domain. This visual comparison is shown in Figure 10. In the spatial filtering visualisation we can quite clearly identify a dominant scale, corresponding to the size of the filter width $\Delta$ indicated by the green bar. This is in contrast to the fine-grained, densely alternating result of the Fourier-based method. Using the definition in Equation (2.4), one can find that the orthogonal expansion using harmonic basis functions encompasses spatially oscillating contributions for the energy flux resulting from the Fourier-based method. Only when spatial averaging is applied, the wavenumber modes that do not form a triad (the well-known triad interactions for spectral energy transfer, see, e.g. Pope 2000; Verma 2019) will average to zero. Following the derivation in Section A, the Fourier-based flux is expected to approach the spatial filtering flux when applying another level of spatial averaging or filtering to the local energy flux obtained from the Fourier-based method. This is showcased numerically by applying a Gaussian filter (using $\Delta = \pi/k$) to the local energy flux results of the Fourier-based method from the leftmost panel of Figure 10. The result is shown in Figure 11. With this filtering in real space, the resulting energy flux distribution from the Fourier-based method closely resembles the result retrieved from the spatial filtering method, as anticipated.

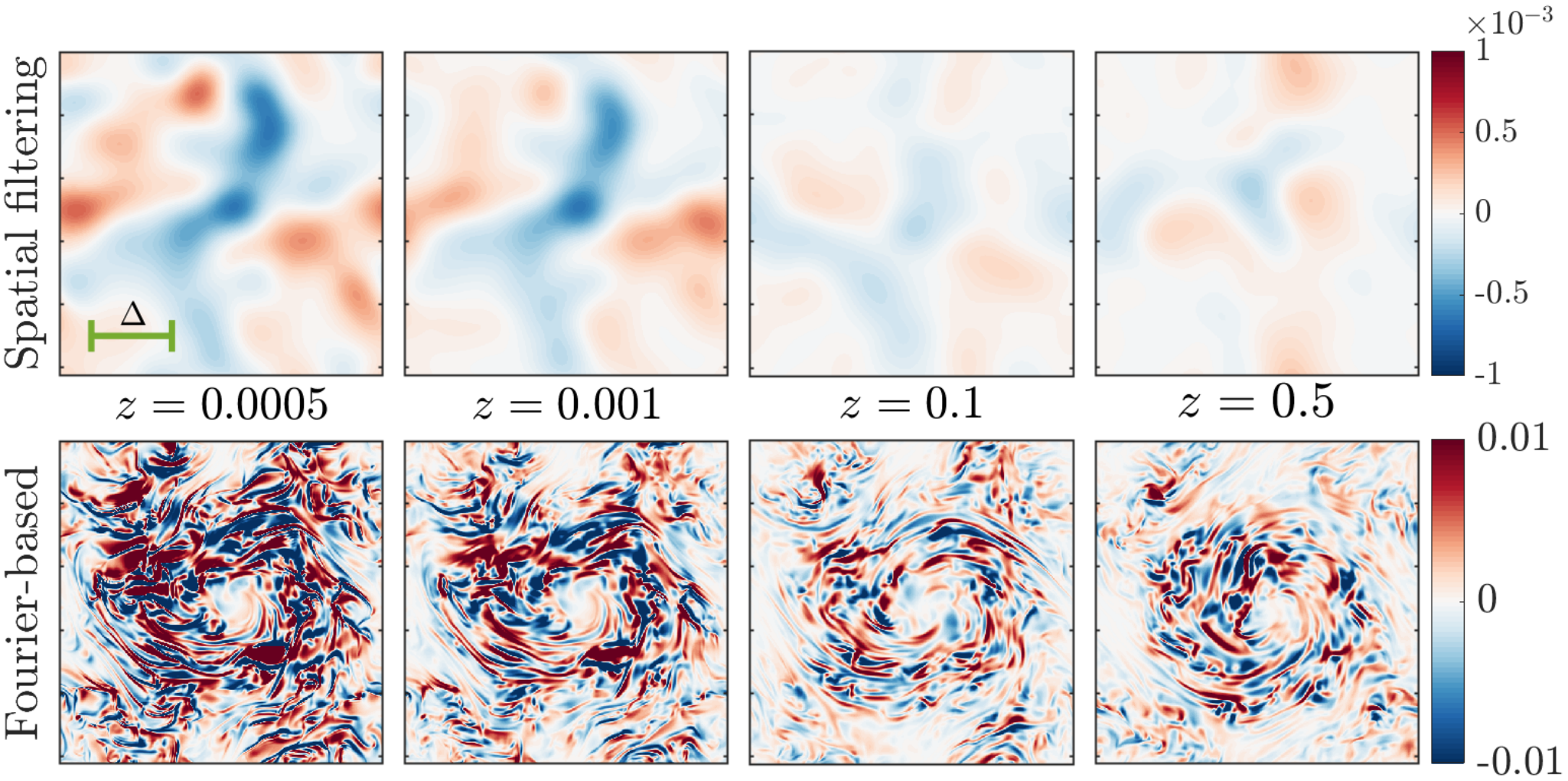

Figure 10. Slices of the*local* scale-by-scale flux of kinetic energy(as defined in Equations (2.4) and (2.8), but *without* spatial averaging $\langle \ldots \rangle$ applied) shown for both analysis methods at four different vertical coordinates for $k = 24.64$. The filter width $\Delta$ that corresponds to this wavenumber is shown in green in the top left panel. The extent of the local difference between the two analysis methods is illustrated here, which can be negated by sufficient spatial and temporal averaging.

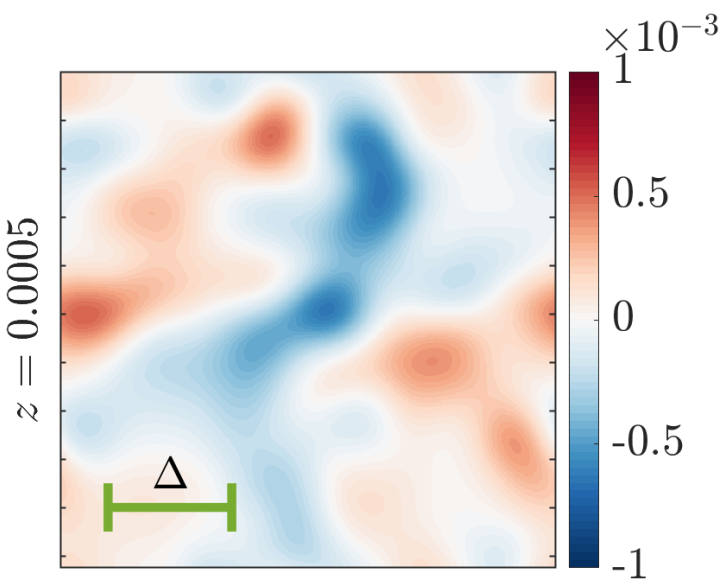

Figure 11. A horizontal slice of Fourier-based energy flux, as seen the leftmost panel of Figure 10, with a Gaussian spatial filter applied to it. This Gaussian filter has a filter width that corresponds to the wavenumber by $\Delta = \pi/k$.

## 4. Rotating Rayleigh-Bénard convection in a cylindrically bounded domain

The results presented in the previous section show a satisfactory agreement between the two analysis methods, revealing the presence of an inverse flux of kinetic energy in our horizontally periodic RRBC simulation. In this section we will consider the case of a fully bounded domain. This is the most challenging domain, imposing further inhomogeneity and aperiodicity, but it is highly relevant to the experimental environment. Additionally, we know that the periodicity in the horizontal directions might enhance the self-correlation of the flow (Frisch 1995), which can affect the energy transport. In this section, we discuss the results of a set of four distinct simulation runs performed on a cylindrically bounded domain; three of the four simulation runs are subjected to various amounts of rotation, while the fourth simulation run is of a nonrotating cylinder. This final simulation run is used as a baseline for the comparison of the kinetic energy flux. Given that cylindrically bounded simulations of RRBC with stress-free boundary conditions have not been widely studied, we have based our choice of input parameters on what was found in Figure 2, together with the same stress-free boundary conditions to enable the possible generation of large flow structures. The Rayleigh number is set to $Ra = 1 \times 10^9$ for all cylinder simulations, while the Ekman number changes between the simulations. The Ekman numbers for the set of cylinder simulation runs consist of $Ek = 1 \times 10^{-6}$, $Ek = 3 \times 10^{-6}$, $Ek = 1 \times 10^{-5}$, for the rotating cylinders, and $Ek = 1 \times 10^{10}$ for the essentially non-rotating cylinder. These cylinder simulation runs are also shown in the regime diagram in Figure 2. Based on the choice of Rayleigh number and the smallest active scales that need to be resolved, a resolution of $(n_\theta, n_r, n_z) = (1152, 384, 768)$ was chosen for each of these simulation runs. With this choice of resolution and aspect ratio, a minimum and maximum grid spacing of $\eta_k$, $\eta_k/50$ was achieved in the vertical coordinate direction. In the horizontal coordinate directions, a minimum and maximum radial grid spacing of $0.28\eta_k$ and $0.57\eta_k$ were achieved, together with a minimum and maximum azimuthal grid spacing of $\eta_k/300$ and $1.0\eta_k$.

The addition of side walls to the simulation domain results in the formation of boundary effects in this region. In some cases, this behaviour can become quite dominant, owing to the formation of a vigorously convective wall mode near the sidewall (Favier & Knobloch 2020; de Wit *et al.* 2020; Zhang *et al.* 2020, 2021; de Wit *et al.* 2023; Zhang *et al.* 2024). The aspect ratio of the cylindrical domain is set to $\Gamma = 0.7$. This aspect ratio is chosen to be slightly larger compared to the aspect ratio of the horizontally periodic simulation to let the bulk of the flow be less influenced by the effect of the side walls. While this side wall region presents an interesting topic for study, our focus lies on the flow in the inner bulk region of the cylinder. To omit the wall region of the cylinder and to enable the implementation of fast Fourier transformation (FFT) on the cylindrical domain, we linearly interpolate the cylindrical velocity field onto a Cartesian grid. From the resulting velocity field, which has a horizontal resolution of $(768, 768)$, we select a square with sides of 512 grid points centred on the cylinder axis. This square thus spans $2/3$ of the diameter $D$ of the cylindrical domain. The edges of this rectangular velocity field are smoothed to zero using a squared sine function over 10 grid points and we pad the final velocity field with zeros in the horizontal directions to avoid effects of implied periodicity when using Fourier transforms. The final resolution of the Cartesian velocity fields that are analysed is $(n_x, n_y, n_z) = (512, 512, 768)$. Given our area of interest we show a quantitative comparison between the spectra of kinetic energy $\hat{E}(k)$ and enstrophy $\hat{\mathcal{E}}(k)$ of the original and windowed inscribed-square velocity fields in Figure 12. From this figure, we can see that the windowing of the simulation domain retains most of the kinetic energy and enstrophy on the dynamically active scales in the inertial range. There is a drop-off in the dissipation range at large $k$, but this range carries only a small proportion of the total energy or enstrophy.

### 4.1. *Flow characterisation*

We begin by characterising the flow with the energy spectra and a visualisation of the flow using the vertical vorticity. In Figure 13 we show the energy spectra for each of the cylinder simulation runs. The lowering of the Ekman number increases the dominance of rotation in the system and thereby decreases the fraction of vertical kinetic energy compared to the horizontal contribution. These energy spectra give an indication of the relative isotropy of the bulk of the flow in each of the cylinders, with cylinder run #1 being the least isotropic and the nonrotating cylinder run the most.

Figure 12. Comparison of spectra of (a) kinetic energy $\hat{E}(k)$ and (b) enstrophy $\hat{\mathcal{E}}(k)$, resulting from an instantaneous velocity field of the full cylinder cross-section and the windowed inscribed square. Also shown are the characteristic wavenumbers of the full cross-section and the inscribed square in black and red dashed lines, respectively.

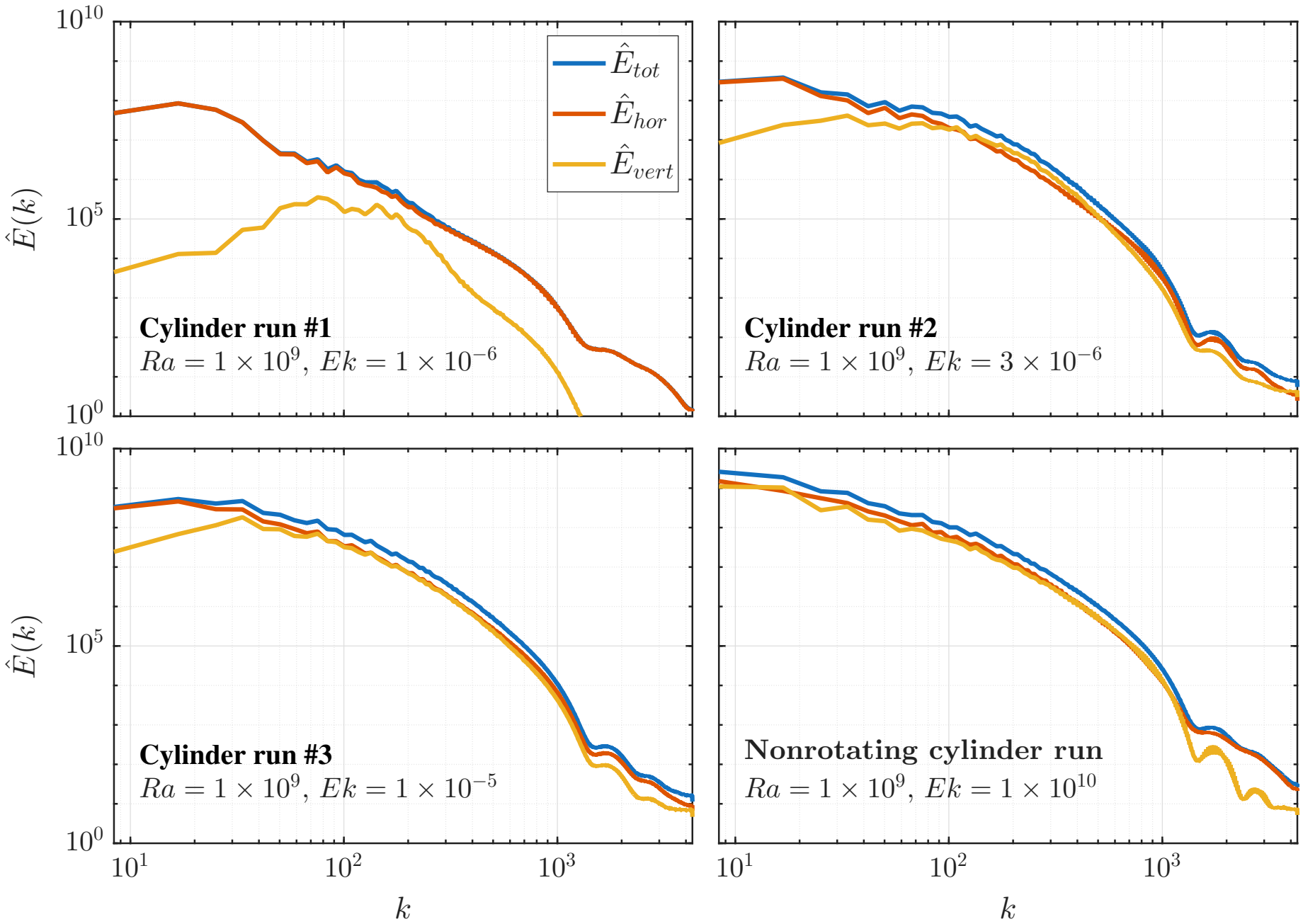

Figure 13. The kinetic energy spectra ($\hat{E}_{tot}$), split out in horizontal ($\hat{E}_{hor.}$) and vertical ($\hat{E}_{vert.}$) components. These represent an instantaneous moment in time. The energy spectra are averaged over a rectangular section from the bulk of the flow in the cylinder.

In Figure 14 a horizontal slice of the cylinders displaying vertical vorticity is shown at two vertical coordinate points to illustrate the flow for each of the cylinder runs. One slice is positioned right above the bottom plate at $z = 0.05$, while the other is at mid-height, $z = 0.5$. These visualisations confirm what we observed in Figure 13: Cylinder run #1 seems to have the most prevalent horizontal structure, with this effect becoming less emphasised as the Ekman number increases. We can clearly identify the prominent effects of rotation on the overall flow structure.

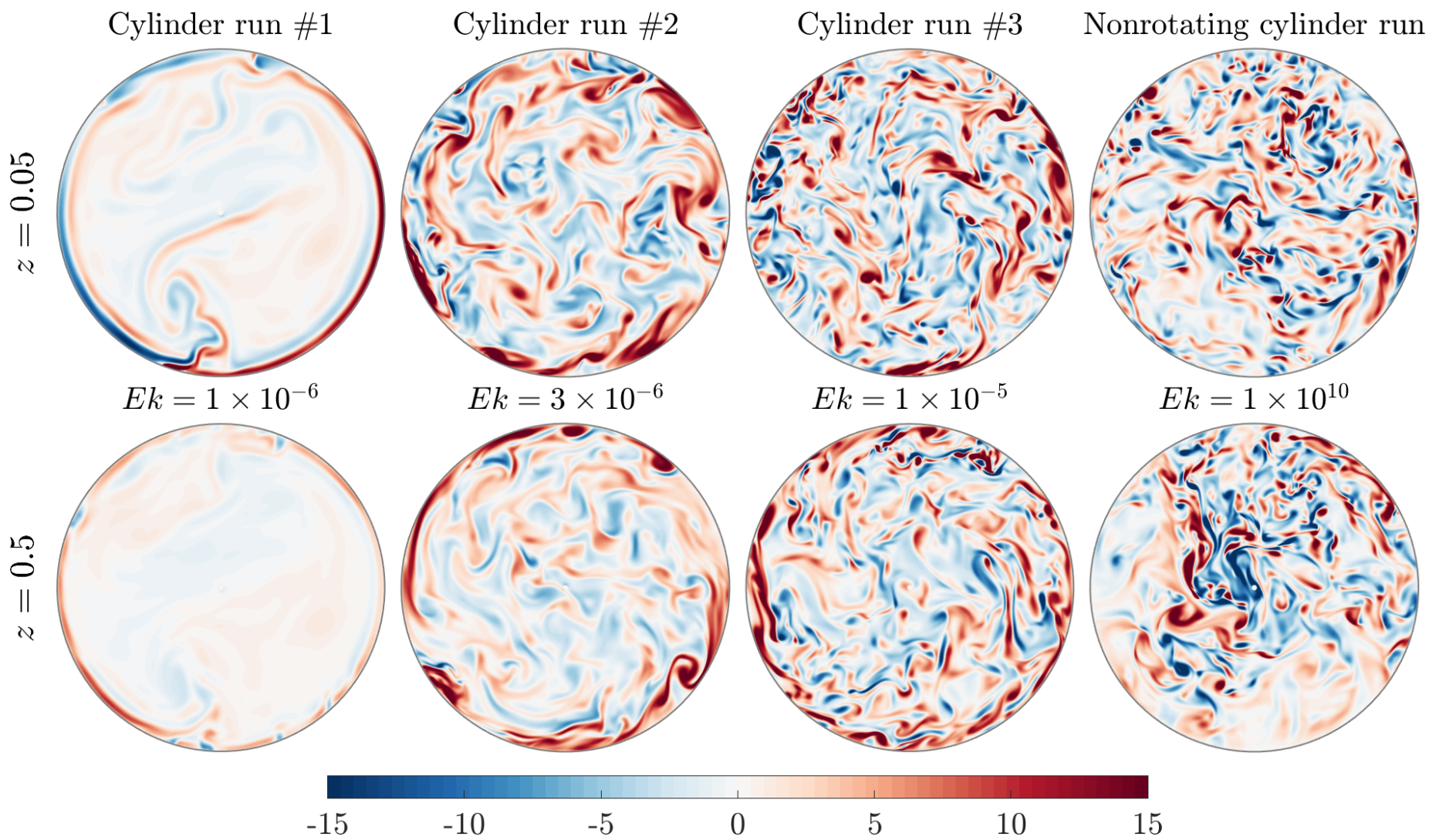

Figure 14. The vertical vorticity $\omega_z$ plotted in horizontal cross-sections of the cylinder runs at $z = 0.5$ and $z = 0.05$.

### 4.2. *The scale-by-scale kinetic energy flux*

As was done with the horizontally periodic simulation, we create a temporal average of the energy flux. For each of the cylinder runs, we use a series of ten instantaneous velocity fields to create this temporal average. The spacing between the velocity fields is kept at $t = 5$. The energy flux resulting from this temporal averaging, including the 95% confidence interval, is shown in Figure 15. In these four windows we should first notice the level of agreement between the two analysis methods. This is similar to what was seen for the horizontally periodic simulation and shows that the agreement holds up, even under the constraints of a fully bounded simulation domain. To give a more in-depth interpretation of these resulting energy fluxes, we start by considering the nonrotating cylinder. For this simulation run, there is no expectation of an inverse flux of energy, which means that a purely positive $\Pi$ is expected. This is consistent with our findings in Figure 15. Switching on rotating and moving to a decreasing Ekman number, we observe that the energy flux decreases and eventually displays a range where it is negative. Although there is definitely a negative range of $\Pi$ seen in runs #1 and #2, the values of the energy flux are several orders of magnitude smaller compared to the nonrotating run. With increasing rotation (smaller $Ek$), the magnitude of the energy flux goes down considerably. This is a nice example of the effect the increase in rotation rate has on the flow in the system, showing how the Coriolis term has a stabilizing, suppressive effect on the convection. This is in line with the findings of earlier studies on the effects of rotation on these types of rotating turbulent flows (Alexakis & Biferale 2018).

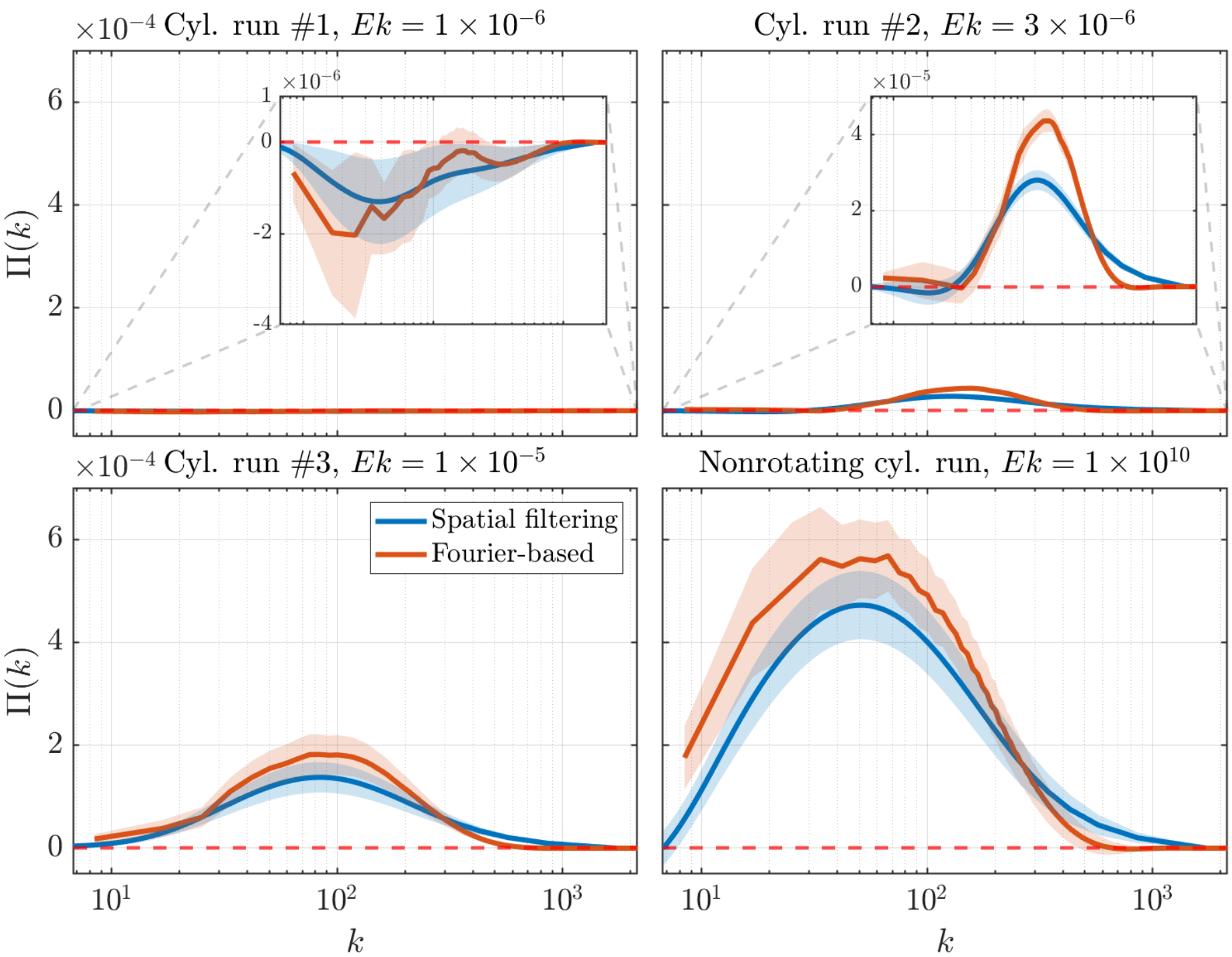

Figure 15. The scale-by-scale kinetic energy flux averaged over a series of instantaneous velocity fields for each of the cylinder simulations. The shaded areas around the curves indicate the 95% confidence interval of the temporal averages. The red dashed line at $\Pi = 0$ represent the boundary between the forward and inverse cascade of energy. The four panels employ the same vertical scale for ease of comparison; insets in the top two panels show these curves on a more appropriate vertical scale.

From the results shown for each of the cylinder simulations, we can identify two distinct conclusions. First, we observe a level of agreement between the two analysis methods similar to that in Section 3. Nonetheless, it should be stated that this agreement is less pronounced, compared to the horizontally periodic case. This effect can be attributed mostly to the additional steps needed to be able to process the velocity data before calculating the Fourier-based energy flux. Nonetheless, the satisfactory agreement is a very promising result which shows that even in this fully bounded domain, a trustworthy computation of the energy flux can be obtained. Second, there is evidence of an inverse cascade for rapidly rotating confined systems.

#### 4.2.1. *Vertical distribution of kinetic energy flux*

As was done with the horizontally periodic domain, we will now take a small step back and try to interpret the behaviour of the scale-by-scale kinetic energy flux throughout the simulation domain. We again do this by averaging the energy flux in horizontal planes and plotting the result as a function of the vertical coordinate. The goal here is to gain insight into how the energy flux depends on the specific location within the domain and how this varies between the different cylinder cases. For simplicity, we limit ourselves to the comparison of only a single wavenumber value across the cylinder cases. For this purpose, a value of $k = 34$ was chosen, based on observations from Figure 15. This wavenumber corresponds to a wavelength that is equivalent to roughly one-fourth of the cylinder diameter. The distribution along the vertical coordinate of the energy flux for the four cylinders is shown in Figure 16.

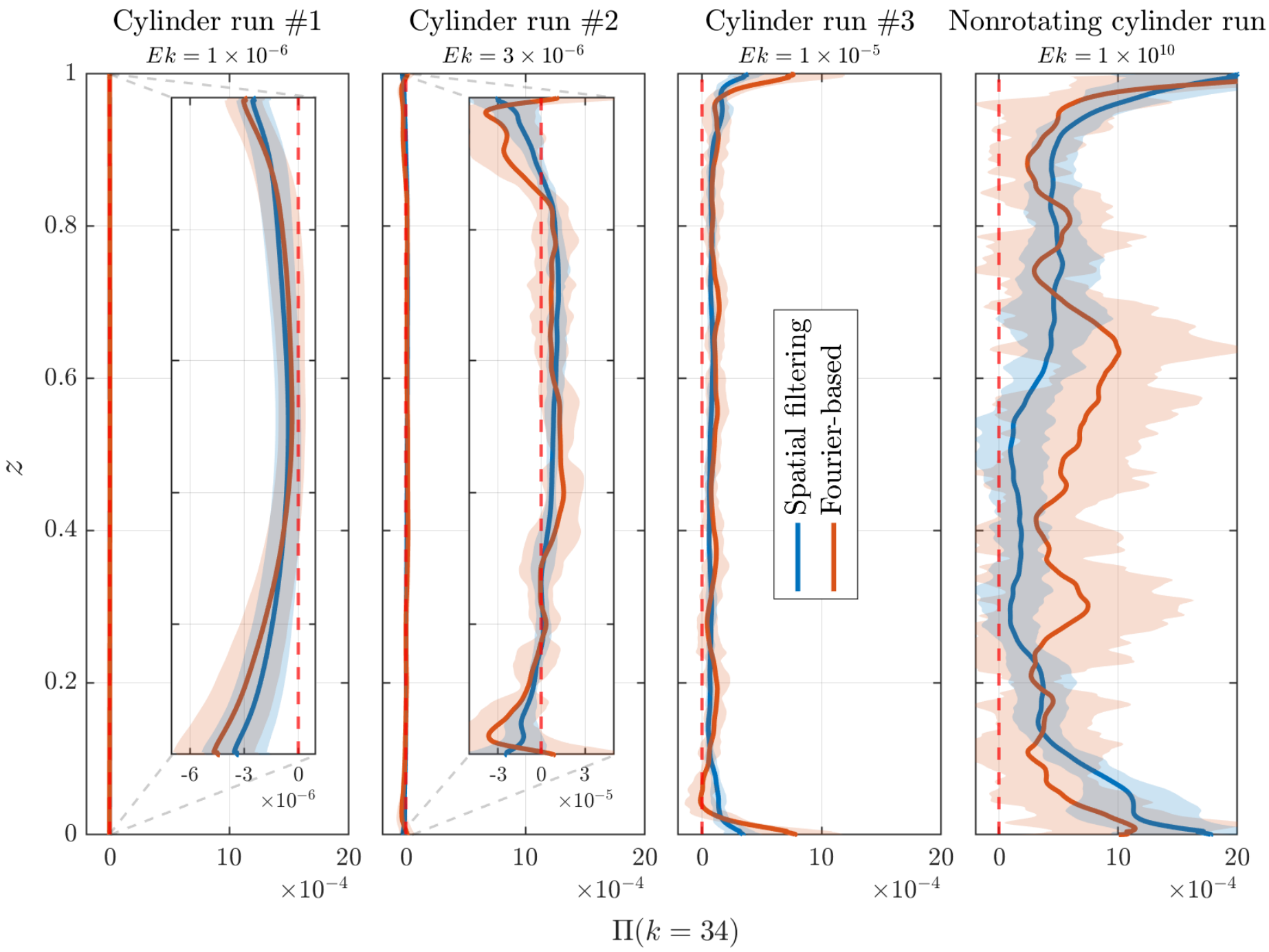

Figure 16. The horizontally averaged scale-by-scale kinetic energy flux plotted against the vertical coordinate for each of the four cylinder simulations runs at $k = 34$. The shaded areas around the curves indicate the 95% confidence interval over the time series. The red dashed line at $\Pi = 0$ represent the boundary between the forward and inverse cascade of energy.

This specific wavenumber was chosen because it best illustrates the dual behaviour of the energy flux that was also observed for the horizontally periodic simulation at the smaller wavenumber values. This dual behaviour means that the bulk of the flow in the middle region of the domain is dominated by the forward cascade ($\Pi > 0$), while in the outer regions ($0 < z < 0.2$ and $0.8 < z < 1$, approximately), the inverse cascade ($\Pi < 0$) is most prevalent. In this figure, we observe the same trend as in Figure 15: The simulation runs with the highest rotation rate possess the clearest inverse energy flux, but the absolute magnitude of this inverse flux is only marginal relative to the forward flux of the runs with a lower rotation rate. Directly near the bounding plates, we see that the energy flux becomes strongly positive. This effect is mainly seen in the two rightmost panels of Figure 16, although cylinder run #2 also shows this in the Fourier-based curve. This effect is similar to what was seen in Figure 9 for the horizontally periodic domain. This outermost region with positive energy flux coincides with the thermal boundary layer of the system, both in the horizontally periodic simulation and in these cylindrically bounded simulations. In Figure 16 we can see that the outer region of the domain is where most energy is injected in the large scales. Finally, comparing the results of the cylinder domain of Figure 16 with the horizontally periodic results of Figure 9, we observe that cylinder run #2 has an energy flux distribution that resembles the horizontally periodic simulation of Figure 9 the most.

## 5. Conclusions

The overarching goal of this paper was to directly compare two distinct analysis methods that quantify the rate at which kinetic energy is transported across different spatial scales.

The secondary objective was to identify and gain insight into the behaviour of this kinetic energy flux in simulations of rotating Rayleigh-Bénard convection in two different geometries. The application of the analysis methods is done with some modifications due to the discretization of the velocity fields, the application to a nonuniformly spaced grid and even due to the application to a bounded cylindrical simulation domain.

The application of both methods leads to an energy flux that is strongly fluctuating in space and time. Spatial and temporal averaging of the energy flux showed that this fluctuating behaviour results in a mean value that shows decent agreement between the analysis methods in most cases. We also show that the inverse cascade of energy is predominantly taking place in regions near the bottom and top plates. This points to a merger of like signed vortical plumes (Julien *et al.* 1996, 2012) as a physical mechanism responsible for the inverse energy transfer. The large-scale vortex gathers all the energy and grows to the largest horizontal scale available in the periodic domain and saturates. Consequently, the vertical middle region of the flow was found to be dominated by a forward cascade of energy.

With the cylindrically bounded simulation domain, we want to bridge the gap to the experimental studies where the flow is constrained by bounding walls on all sides. Even in this more challenging domain, a satisfactory degree of agreement between both analysis methods was found, after careful preprocessing steps. Furthermore, in the cylindrical simulation runs, we were also able to identify most of the characteristics of the analysis methods that were found in the treatment of the horizontally periodic simulation run.

The level of agreement between the Fourier-based method and the spatial filtering method that we found is reassuring and should enable the application of either method to numerical and experimental works in a broader context. The biggest distinction between the analysis methods that was identified lies with the implementation of the length scales at which the energy flux is calculated. For the spatial filtering method, this is an arbitrary choice, making this method more versatile. This is in contrast to the Fourier-based method, where the wavenumbers at which the flux is calculated are directly correlated to the spacing of the grid, and with that the chosen resolution. We find furthermore that the spatial filtering based methods tends to yield smaller error bars. We finally point out that a filtering-based method can more easily be adapted to complex geometries than the Fourier-based method considered here.

We plan to incorporate these energy flux calculations into the code for an immediate run-time averaging rather than post-processing from individual velocity fields. This will give a time-averaged output with a much higher degree of convergence. Such an implementation will also greatly increase the level of confidence in the outcome, as now the post-processing was a limiting factor for the extent to which the energy flux could be averaged in the temporal domain.

**Funding.** This publication is part of the project 'Universal critical transitions in constrained turbulent flows' with file number VI.C.232.026 of the research programme NWO-Vici which is financed by the Dutch Research Council (NWO). We are grateful for the support of NWO for the use of supercomputer facilities (Snellius) under Grants No. 2023.013 and 2025.011.

**Declaration of Interests.** The authors report no conflict of interest.

## Appendix A. Flux formula closing

In this appendix, we show analytically under what conditions the expressions Equation (2.4) and Equation (2.8) become equivalent. These arguments are based on the definitions of the energy flux in Frisch (1995); Pope (2000); Sagaut (2005) and the analytical work done by Aluie & Eyink (2009).

### A.1. *Fourier-based flux*

We start from the expression for the Fourier-based flux

$$\Pi^{(\text{Fourier})}(k) \equiv \boldsymbol{u}_k^< \cdot (\boldsymbol{u} \cdot \boldsymbol{\nabla})\boldsymbol{u}_k^> . \tag{A 1}$$

We now write this with the filtering operator $\bar{\cdot}$ as

$$\Pi^{(\text{Fourier})} \equiv \bar{\boldsymbol{u}} \cdot (\boldsymbol{u} \cdot \boldsymbol{\nabla})(\boldsymbol{u} - \bar{\boldsymbol{u}}). \tag{A 2}$$

We now consider the filtered version of this flux

$$\overline{\Pi}^{(\text{Fourier})} = \overline{\bar{\boldsymbol{u}} \cdot (\boldsymbol{u} \cdot \boldsymbol{\nabla})(\boldsymbol{u} - \bar{\boldsymbol{u}})}. \tag{A 3}$$

If the filtering operator is a Reynolds operator, we can use the property that $\bar{\bar{\phi}} = \bar{\phi}$ (Sagaut 2005). The Fourier filter is an exact Reynolds operator, but the Gaussian filter only approximately so. Steps in the derivation that rely on the properties that are exclusive to Reynolds operators will be denoted by $\hat{=}$.

We can simplify this to

$$\begin{aligned}
\overline{\Pi}^{(\text{Fourier})} &\hat{=} \bar{\boldsymbol{u}} \cdot \overline{(\boldsymbol{u} \cdot \boldsymbol{\nabla})(\boldsymbol{u} - \bar{\boldsymbol{u}})}, \\
&= \bar{\boldsymbol{u}} \cdot (\overline{(\boldsymbol{u} \cdot \boldsymbol{\nabla})\boldsymbol{u} - (\boldsymbol{u} \cdot \boldsymbol{\nabla})\bar{\boldsymbol{u}}}), \\
&= \bar{\boldsymbol{u}} \cdot (\overline{(\boldsymbol{u} \cdot \boldsymbol{\nabla})\boldsymbol{u}} - \overline{(\boldsymbol{u} \cdot \boldsymbol{\nabla})\bar{\boldsymbol{u}}}), \\
&= \bar{\boldsymbol{u}} \cdot \overline{(\boldsymbol{u} \cdot \boldsymbol{\nabla})\boldsymbol{u}} - \bar{\boldsymbol{u}} \cdot \overline{(\boldsymbol{u} \cdot \boldsymbol{\nabla})\bar{\boldsymbol{u}}}, \\
&\hat{=} \bar{\boldsymbol{u}} \cdot \overline{(\boldsymbol{u} \cdot \boldsymbol{\nabla})\boldsymbol{u}} - \bar{\boldsymbol{u}} \cdot (\bar{\boldsymbol{u}} \cdot \boldsymbol{\nabla})\bar{\boldsymbol{u}}.
\end{aligned} \tag{A 4}$$

In index notation, this becomes

$$\overline{\Pi}^{(\text{Fourier})} = \overline{u_i}\,\overline{u_j \frac{\partial u_i}{\partial x_j}} - \overline{u_i}\,\overline{u_j}\frac{\partial \overline{u_i}}{\partial x_j}. \tag{A 5}$$

### A.2. *Filtering-based flux*

The filtering-based flux is defined as

$$\Pi^{(\text{filter})} \equiv -\tau_{ij}\bar{S}_{ij}, \tag{A 6}$$

with

$$\tau_{ij} \equiv \overline{u_i u_j} - \overline{u_i}\,\overline{u_j}, \qquad \bar{S}_{ij} \equiv \frac{1}{2}\left(\frac{\partial \overline{u_i}}{\partial x_j} + \frac{\partial \overline{u_j}}{\partial x_i}\right). \tag{A 7}$$

This expands to

$$\Pi^{(\text{filter})} = -\tfrac{1}{2}\left(\overline{u_i u_j} - \overline{u_i}\,\overline{u_j}\right)\left(\frac{\partial \overline{u_i}}{\partial x_j} + \frac{\partial \overline{u_j}}{\partial x_i}\right), \tag{A 8a}$$

$$= -\tfrac{1}{2}\left(\overline{u_i u_j}\frac{\partial \overline{u_i}}{\partial x_j} + \overline{u_i u_j}\frac{\partial \overline{u_j}}{\partial x_i} - \overline{u_i}\,\overline{u_j}\frac{\partial \overline{u_i}}{\partial x_j} - \overline{u_i}\,\overline{u_j}\frac{\partial \overline{u_j}}{\partial x_i}\right). \tag{A 8b}$$

Now the first two terms and the second two terms are identical under exchange of dummy indices $i$ and $j$, such that

$$\Pi^{(\text{filter})} = -\overline{u_i u_j}\frac{\partial \overline{u_i}}{\partial x_j} + \overline{u_i}\,\overline{u_j}\frac{\partial \overline{u_i}}{\partial x_j} \equiv \Pi_A + \Pi_B. \tag{A 9}$$

The first term can be rewritten using partial integration to find that

$$\Pi_A = -\overline{u_i u_j}\frac{\partial \overline{u_i}}{\partial x_j} = \overline{u_i}\frac{\partial \overline{u_i u_j}}{\partial x_j}, \tag{A 10}$$

where the boundary term vanished due to periodic (or impermeable) boundary conditions. Then, using commutivity between the filter and differential operator, we obtain

$$\Pi_A = \overline{u}_i \overline{\frac{\partial u_i u_j}{\partial x_j}}, \tag{A 11}$$

and employing the incompressibility condition $\partial u_j / \partial x_j = 0$, we find

$$\Pi_A = \overline{u}_i \overline{u_j \frac{\partial u_i}{\partial x_j}}. \tag{A 12}$$

Substituting this back into the flux expression finally yields

$$\Pi^{(\text{filter})} = \Pi_A + \Pi_B = \overline{u}_i \overline{u_j \frac{\partial u_i}{\partial x_j}} + \overline{u}_i \overline{u}_j \frac{\partial \overline{u}_i}{\partial x_j}. \tag{A 13}$$

### A.3. *Equivalence*

Comparing Equation (A 5) and Equation (A 13), we find that these expressions are equivalent, apart from the term $\overline{u}_i \overline{u}_j \frac{\partial \overline{u}_i}{\partial x_j}$. This term captures the sweeping by the filtered field (Aluie & Eyink 2009). The term can be rewritten (using the incompressibility condition) as

$$\overline{u}_i \overline{u}_j \frac{\partial \overline{u}_i}{\partial x_j} = \frac{\partial \left( \frac{1}{2} \overline{u}_i^2 \overline{u}_j \right)}{\partial x_j}, \tag{A 14}$$

which is a divergence term, so using the periodic (or impermeable) boundary condition, it can be seen that this term integrates to zero, and thus has no net contributions to the spatial average.

In conclusion, for filter operators that are Reynolds operators such as the Fourier filter, flux expressions Equation (A 3) and Equation (A 6) are equivalent when integrated over the full domain. Locally, they are different by a term $2\overline{u}_i \overline{u}_j \frac{\partial \overline{u}_i}{\partial x_j}$ called the sweeping term. For larger length scales, the filter operation approaches domain averaging, so the sweeping term approaches zero.

The Gaussian filter is not a Reynolds filter, so only the expression in Equation (A 6) correctly computes the energy flux in that case, while Equation (A 3) does not.

## Appendix B. Spatial filtering shape comparison

In order to justify the choice to use a Gaussian filter for the spatial filtering approach, and to confirm that the presented results are not an effect of the chosen filter shape, we use this appendix to compare the results of the spatial filtering method when using a Gaussian (Equation (2.6)) or a box filter

$$G_{\text{box}}(\boldsymbol{x}) \equiv \begin{cases} \displaystyle\prod_i \frac{1}{\Delta} & \text{if } x_i \leq \dfrac{\Delta}{2}, \\ 0 & \text{otherwise,} \end{cases} \tag{B 1}$$

where the subscript $i$ denotes the coordinate direction. These filter shapes share a similarity in the fact that they are mostly compact in real space and thus the filtered velocity fields are ostensibly comparable. The main difference between these filter shapes is their behaviour in spectral space. Where the Gaussian filter is still mostly compact there, the box-shaped filter follows a cardinal sine function:

$$\widehat{G}_{\text{box}}(k) \equiv \text{sinc}\left(\frac{k\Delta}{2}\right) = \frac{\sin\left(\frac{k\Delta}{2}\right)}{\frac{k\Delta}{2}}. \tag{B 2}$$

In Figure 17, we show the direct comparison of the domain averaged kinetic energy flux calculated using the spatial filtering method for a single snapshot, where the only difference is the shape of the applied spatial filter. This figure clearly indicates that the spatial filtering results are not very sensitive to the choice of the filter shape.

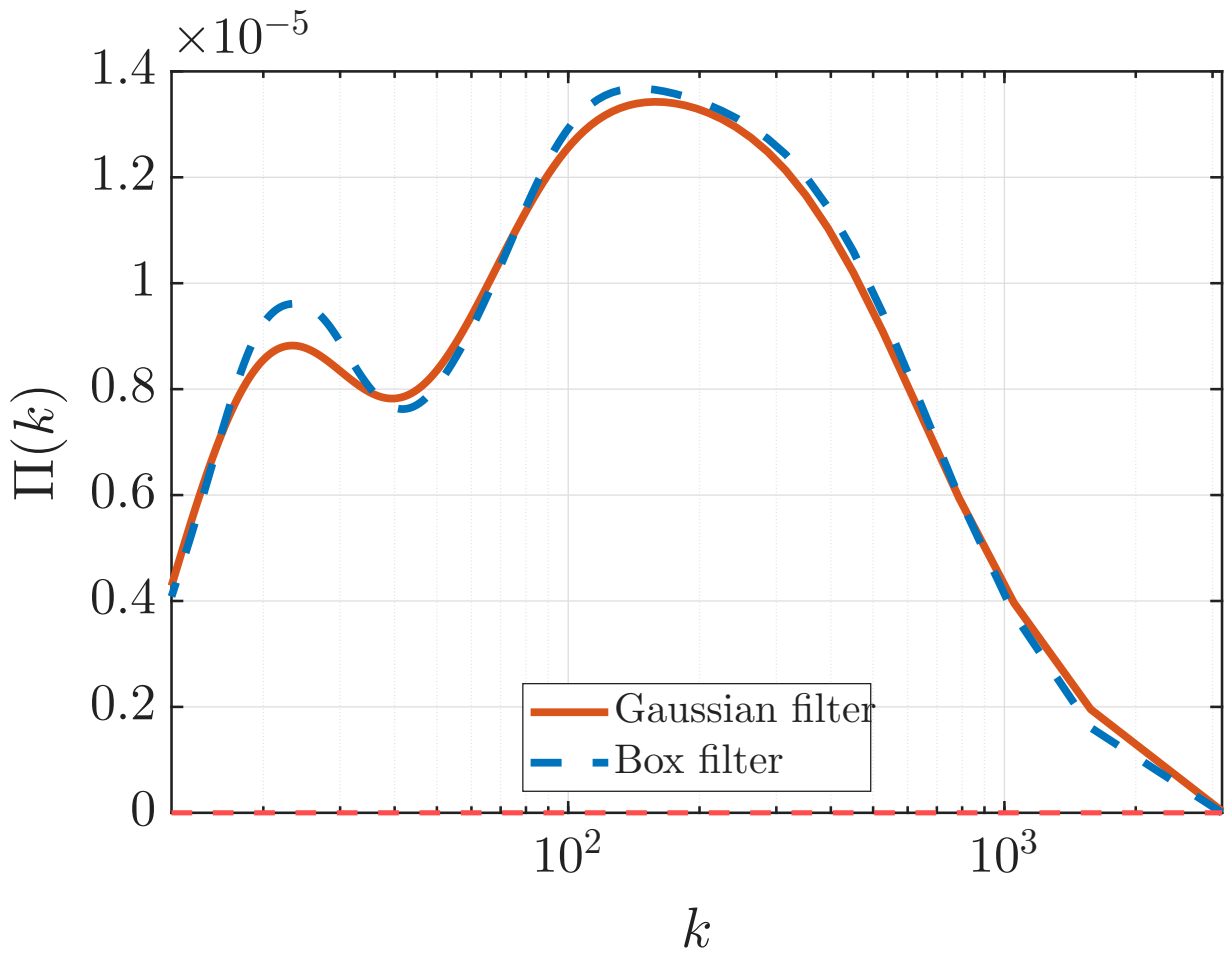

Figure 17. The kinetic energy flux of an instantaneous velocity field of the horizontally periodic RRBC simulation. This energy flux is calculated using the spatial filtering method, using a Gaussian and a box-shaped filter.